\documentclass[preprint,pre,showpacs,preprintnumbers,amsmath,amssymb]{revtex4}

\usepackage{graphicx}% Include figure files
\usepackage{dcolumn}% Align table columns on decimal point
\usepackage{bm}% bold math

\begin{document}

\preprint{PREPRINT}

\title{Generalized Lattice-Boltzmann Equation with Forcing Term for Computation of Wall-Bounded Turbulent Flows}% Force line breaks with \\

\author{Kannan N. Premnath}
\email{nandha@metah.com}

\affiliation{Department of Chemical Engineering, University of
California, Santa Barbara, Santa Barbara, CA 93106\\}

\affiliation{ MetaHeuristics LLC, 3944 State Street, Suite 350,
Santa Barbara, CA 93105}

\author{Martin J. Pattison}
\email{martin@metah.com}

\affiliation{ MetaHeuristics LLC, 3944 State Street, Suite 350,
Santa Barbara, CA 93105}

\author{Sanjoy Banerjee}
\email{banerjee@engineering.ucsb.edu}

\affiliation{Department of Chemical Engineering \\ Department of
Mechanical Engineering \\ Bren School of Environmental Science and
Management\\ University of California, Santa Barbara, Santa Barbara,
CA 93106\\}

%\affiliation{
%Department of Chemical Engineering\\
%University of California, Santa Barbara \\
%Santa Barbara, CA 93106}
%\affiliation{
%MetaHeuristics LLC\\
%3944 State Street, Suite 350\\
%Santa Barbara, CA 93105}

%\author{Kannan N. Premnath}
%\email{nandha@metah.com}
%% \homepage{http://www.Second.institution.edu/~Charlie.Author}
%\affiliation{
%Department of Chemical Engineering\\
%University of California, Santa Barbara\\
%Santa Barbara, CA 93106
%%This line break forced% with \\
%}%
%
%%\altaffiliation[Also at ]{MetaHeuristics LLC.}%Lines break automatically or can be forced with \\
%%%\author{Kannan N. Premnath}%
%%\email{nandha@metah.com}
%
%\author{Martin J. Pattison}
%\email{martin@metah.com}
% %\altaffiliation[Also at ]{Physics Department, XYZ University.}%Lines break automatically or can be forced with \\
%%\author{Kannan N. Premnath}%
%% \email{Second.Author@institution.edu}
%\affiliation{%
%MetaHeuristics LLC\\
%3944 State Street, Suite 350,\\
%Santa Barbara, CA 93105
%%This line break forced with \textbackslash\textbackslash
%}%
%
%\author{Sanjoy Banerjee}
%\email{banerjee@engineering.ucsb.edu}
% %\altaffiliation[Also at ]{Physics Department, XYZ University.}%Lines break automatically or can be forced with \\
%%\author{Kannan N. Premnath}%
%% \email{Second.Author@institution.edu}
%\affiliation{
%Department of Chemical Engineering\\
%University of California, Santa Barbara\\
%Santa Barbara, CA 93106
%%This line break forced% with \\
%}%

\date{\today}% It is always \today, today,

\begin{abstract}
In this paper, we present a framework based on the generalized lattice-Boltzmann
equation (GLBE) using multiple relaxation times with forcing term for eddy capturing
simulation of wall bounded turbulent flows. Due to its flexibility in using
disparate relaxation times, the GLBE is well suited to maintaining numerical
stability on coarser grids and in obtaining improved solution fidelity of near-wall
turbulent fluctuations. The subgrid scale (SGS) turbulence effects are represented
by the standard Smagorinsky eddy-viscosity model, which is modified by using the
van Driest wall-damping function to account for reduction of turbulent length scales
near walls. In order to be able to simulate a wider class of problems, we introduce
forcing terms, which can represent the effects of general non-uniform forms of forces,
in the natural moment space of the GLBE. Expressions for the strain rate tensor used
in the SGS model are derived in terms of the non-equilibrium moments of the GLBE to
include such forcing terms, which comprise a generalization of those presented in
a recent work (Yu \emph{et~al.}, Comput. Fluids, $\bf{35}$, 957 (2006)). Variable
resolutions are introduced into this extended GLBE framework through a conservative
multiblock approach. The approach, whose optimized implementation is also discussed,
is assessed for two canonical flow problems bounded by walls, viz., fully-developed
turbulent channel flow at a shear or friction Reynolds number ($\mathrm{Re}$) of $183.6$
based on the channel half-width and three-dimensional (3D) shear-driven flows
in a cubical cavity at a $\mathrm{Re}$ of 12,000 based on the side length of the cavity.
Comparisons of detailed computed near-wall turbulent flow structure, given in
terms of various turbulence statistics, with available data, including those
from direct numerical simulations (DNS) and experiments showed good
agreement. The GLBE approach also exhibited markedly better stability characteristics
and avoided spurious near-wall turbulent fluctuations on coarser grids when compared with
the single-relaxation-time (SRT)-based approach. Moreover, its implementation showed
excellent parallel scalability on a large parallel cluster with over
a thousand processors.
\end{abstract}

\maketitle

\section{\label{sec:intro}Introduction}
The lattice Boltzmann method (LBM), employing minimal discrete
kinetic models to solve fluid mechanics and other physical problems,
has attracted much attention in recent
years~\cite{chen98,succi01,succi02,yu03}. Instead of directly
solving the Navier--Stokes equations (NSE), the LBM involves the
solution of the lattice Boltzmann equation
(LBE)~\cite{mcnamara88,higuera89a,higuera89b,qian92,chen92}, which
describes the evolution of the distribution of particle populations
on a lattice whose collective behavior asymptotically reproduces the
dynamics of fluid flow. More specifically, the lattice, possessing
sufficient rotational and other symmetries, restricts the collisions
and movements of particle populations along discrete directions, as
represented by the LBE, in such a way that in the continuum limit, fluid
flow represented by weakly compressible NSE is recovered. While its origins
lie in the lattice gas cellular automata (LGCA)~\cite{frisch86}, its formal
connection to kinetic theory~\cite{he97a,he97b} has more recently led to
improved physical modeling using the LBM, for example to represent
multiphase flows~\cite{he02}, and greater amenability for numerical
analysis~\cite{junk05}. The attractiveness of the LBM comes from the
simplicity of the stream-and-collide computational procedure,
absence of the need for an elliptic Poisson-type equation for the pressure
field, ease in handling boundary conditions for representation
of complex geometries, and excellence parallel performance due to its
explicit and local nature. As a result, it has found a number of
interesting fluid flow applications~\cite{succi01,yu03,nourgaliev03}.

Representation and computation of turbulence is one of the most
challenging aspects of fluid dynamics~\cite{frisch95,pope00}. In
recent years, significant progress has been made to derive
turbulence models \emph{a priori} from discrete kinetic
theory~\cite{chen98a,ansumali04,chen04}, and turbulence modeling
in the LBM has found much success in practical applications,
for e.g., by Teixeira~\cite{teixeira98} and Chen \emph{et~al.}~\cite{chen03}.
Also, various prior studies have found that LBM is a reliable and accurate
method for direct numerical simulation (DNS) of various benchmark turbulent
flow problems -- see for e.g. Refs.~\cite{martinez94,amati97,amati99,yu05a,yu05b,yu05c,yu06a,lammers06}.

On the one hand, turbulence models in the Reynolds-averaged contexts are generally
required to represent physics over a wide range of scales. While turbulence at small scales
tends to be somewhat more universal, large scale turbulent motions are strongly problem dependent.
Hence, it is unrealistic to expect Reynolds-averaged models to accommodate and represent large-scale
behavior of different classes of turbulent flows in the same manner without resorting to considerable
empiricism. On the other hand, the DNS approach resolves all relevant spatial and temporal scales and
can thus predict all possible fluid motions with high fidelity. However, its computational cost limits
its utility to low Reynolds numbers. Thus, it is often more practical to use large eddy simulations (LES),
where fluid motions with length scales greater than the grid size are computed and the effect of the
unresolved eddies at subgrid scales (SGS) are modeled~\cite{sagaut02}. In this regard, while the use of
simple Smagorinsky model~\cite{smagorinsky63} to represent SGS effects and perform LES using LBM was
proposed some time ago by Hou \emph{et~al}.~\cite{hou96} and Eggels~\cite{eggels96}, it has only more
recently found applications for flows in different configurations and physical conditions -- see for e.g.
Refs.~\cite{derksen99,lu02,krafczyk03,hartmann04,yu05c}.

The effects of particle collisions in the solution of LBE are
generally represented by relaxation-type models. One of the most
common among them is the single-relaxation-time (SRT) model, also
termed as the Bhatnagar-Gross-Krook (BGK) model~\cite{bhatnagar54}.
Owing to its simplicity, the use of the SRT model in the
LBE~\cite{qian92,chen92} has been popular for simulating a variety
of problems, including the computation of turbulent flow problems
mentioned above. It is well known that the SRT model is quite
susceptible to numerical instabilities when it is employed for
simulating high Reynolds number flows~\cite{lallemand00}. In
particular, the lack of proper mechanisms to properly dissipate
unphysical small-scale oscillations arising due to non-hydrodynamic
or kinetic modes in the LBE can often cause numerical
instabilities~\cite{dellar01}. In the case of turbulent flows, and
more specifically in coarser grid eddy-capturing simulations, such
spurious oscillations may interfere with turbulent fluctuations and
can result in loss of accuracy and stability. An important approach
to enhance numerical stability with using SRT models is through the
entropic lattice Boltzmann methods (ELBM), which ensures positivity
of the distribution
functions~\cite{karlin99,ansumali02,ansumali05,boghosian01}. While
being endowed with elegant and desirable physical features, it may
be noted that they have certain computational and physical
limitations, as pointed in Refs.~\cite{wong03,wong07}, and are not
the pursued in this current work.

A more general form of the LBE, sometimes also called the moment
method or the generalized lattice Boltzmann equation (GLBE), is
based on the use of multiple relaxation times (MRT) to represent
collision effects~\cite{dhumieres92}. It is actually a refined form
of the quasi-linear relaxation version of LBE with a collision
matrix~\cite{higuera89a,higuera89b,benzi92}, where collision is carried out
in the moment space. In contrast to the
SRT-LBE, the MRT-LBE or GLBE deals with moments of the distribution
functions, such as momentum and viscous stress directly. This moment
representation provides a natural and convenient way to express
various relaxation processes due to collisions, which often occur at
different time scales. Also, the collision matrix takes a much
reduced form as a diagonal matrix in this moment space. By carefully
choosing and separating different time scales to represent changes
in various hydrodynamic and kinetic modes through a von Neumann
stability analysis of the kinetic equation~\cite{resibois77}, the
numerical stability of the LBE can be significantly
improved~\cite{lallemand00}. The general forms of the MRT models
in two-dimensions (2D) and three-dimensions (3D) are presented by
Lallemand and Luo~\cite{lallemand00} and d'Humi\`{e}res \emph{et~al}.~\cite{dhumieres02},
respectively. Simplified forms of MRT models~\cite{ladd94,ginzburg03,ginzburg05}
and with a different weighted representation of moments~\cite{adhikari05} have also been
introduced to improve boundary conditions and to improve ability to
represent hydrodynamics with thermal fluctuations.
The MRT-LBE has been further extended with the use of additional forcing terms
to simulate complex fluid flows, such as multiphase flows in 2D and 3D by
McCracken and Abraham~\cite{mccracken05} and
Premnath and Abraham~\cite{premnath06}, respectively, and applied to simulate
complex multiphase flow problems with significantly enhanced numerical
stability~\cite{premnath05a,mccracken05a}. More recently, the GLBE
approach~\cite{premnath06} has also been used to simulate complex
magnetohydrodynamic problems with much success~\cite{pattison07}.

In recent years, Yu \emph{et~al}.~\cite{yu06} developed a MRT-LBE for LES of
certain classes of turbulent flows. In particular, they employed the Smagorinsky
SGS model with a constant coefficient, where the local strain rate tensor is given
in terms of non-equilibrium moments. As such, their approach is applicable for
problems without boundary effects, for e.g., free-shear flows and they have indeed
validated it for a turbulent free-jet flow problem. However, the presence of either
stationary or moving boundaries, such as walls or free surfaces, respectively, are
known to strongly affect the turbulence structures, and suitable modifications are
needed to the standard Smagorinsky SGS model for use with the GLBE.
Moreover, in many situations, external forces, such as constant body forces
mimicking pressure gradient in a periodic domain or non-uniform forces
such as Lorentz or Coriolis forces, can drive and/or strongly influence the character
of turbulent flow physics. The effects of these forces can be introduced as forcing terms
in the GLBE. Also, the use of forcing terms representing non-uniform forces provide
a framework to introduce more general forms of SGS Reynolds stress models that are not
based on eddy-viscosity assumption. Moreover, as the scales of turbulent flow vary locally
in general situations, it is important to employ local grid refinement approaches in
conjunction with the MRT-LBE.

Thus, a primary objective of this paper is to develop a framework for LES using the MRT-LBE
with forcing term for wall-bounded flows, in which near-wall turbulence is generally known
to be anisotropic and inhomogeneous in nature. We propose to carry out the forcing term
in the natural moment space of the GLBE so that it is readily amenable for simulating general
forms of non-uniform forces. The computations of the moment-projections of the forcing term
are provided for the three-dimensional, nineteen velocity (D3Q19) model~\cite{qian92}.
To account for the reduction in the turbulent length scale near walls, we employ the van Driest
wall damping function~\cite{vandriest56} in the Smagorinsky SGS model. We derive
expressions for the strain rate tensor used in the SGS model in terms of the non-equilibrium
moments of the GLBE in the presence of forcing terms representing general non-uniform forces
by means of Chapman--Enskog analysis~\cite{chapman64,premnath06}, which is a generalization of those
presented by Yu~\emph{et~al.}~\cite{yu06}. We also briefly discuss an optimized computational
procedure for such an extended GLBE formulation. Moreover, we incorporate variable resolutions
in the GLBE by introducing a conservative local grid refinement approach~\cite{chen06,rohde06}.
While the use of a constant Smagorinsky SGS model is known to have certain limitations
(see, for e.g., Ref.~\cite{sagaut02}), as a first step to model bounded flows as well as for
reasons of computational efficiency, we have employed it in conjunction with the damping function,
which are known to be reasonably accurate for certain wall-bounded flows~\cite{moin82}.
It may be noted that more sophisticated SGS models involving the use of dynamic procedures to
determine the values of the parameters in the SGS models~\cite{germano91} that circumvents some
of the limitations of the constant Smagorinsky SGS model have also been successfully
used in the LBM context recently by Premnath \emph{et~al.}~\cite{premnath08a}. Another important
recent development is an inertial-consistent Smagorinsky SGS model proposed for use with
the LBM by Dong \emph{et~al.}~\cite{dong08}.

Another objective of this paper is to perform systematic studies for assessment of accuracy
and gains in numerical stability using the LES framework described above for a set of canonical
wall-generated flow turbulence problems. In particular, we evaluate the GLBE in detail for two
problems viz., fully-developed turbulent channel flow at a shear or friction Reynolds number $\mathrm{Re}$
of $183.6$ based on channel half-height and 3D driven cavity flows at a $\mathrm{Re}$ of $12,000$ based
on cavity side width. The benchmark problems involve complex features of wall-bounded turbulent
flows, and extensive prior data, including those from DNS and experiments are available to compare
and assess the results of the detailed structure of turbulence statistics obtained using the GLBE
computations. We also study the gains in numerical stability when the GLBE is used in lieu of the
SRT-LBE for such complex anisotropic and inhomogeneous turbulent flows as well as the parallel
scalability of its implementation on a massively parallel cluster.

This paper is organized as follows. In Section~\ref{sec:glbe}, we
discuss the development of the generalized lattice-Boltzmann equation
(GLBE) with forcing term. Section~\ref{sec:turbmodel} presents the
subgrid scale model for wall-bounded turbulent flows used in this
work. Details of the computational procedure of GLBE and its
optimization are provided in Sec.~\ref{sec:compoptimize}.
The simulation results, accuracy and stability of two canonical problems,
viz., fully-developed turbulent channel flow and 3D cubical cavity flow
are discussed in Secs.~\ref{sec:turbchannelflow} and~\ref{sec:3dcavityflow},
respectively. Finally, the summary and conclusions are presented in
Sec.~\ref{sec:summary}. More elaborate details of the approach used
in this work are presented in various appendices.

\section{\label{sec:glbe}Generalized Lattice Boltzmann Equation with Forcing Term}
\begin{figure}
\includegraphics[width = 80mm,viewport=185 150 540
460,clip]{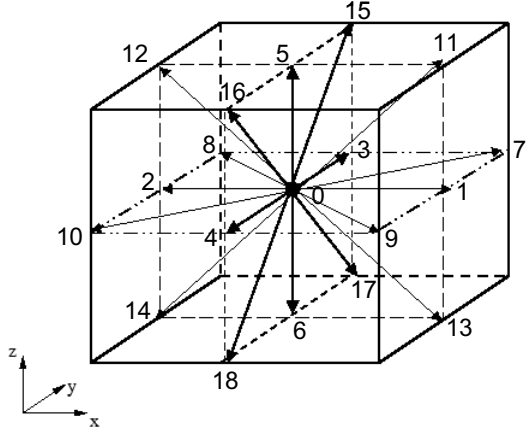}% Here is how to import EPS art
\caption{\label{fig:d3q19} Schematic illustration of the
three-dimensional, nineteen velocity (D3Q19) model.}
\end{figure}

The computational approach for turbulent flows based on the solution
of the GLBE is a recent version of the LBM. The GLBE consists of the evolution equation of the
distribution function $f_{\alpha}$ of particle populations as they
move and collide on a lattice and is given
by~\cite{dhumieres02,premnath06}
\begin{equation}
f_{\alpha}(\overrightarrow{x}+\overrightarrow{e_{\alpha}}\delta_t,t+\delta_t)-f_{\alpha}(\overrightarrow{x},t)=
-\sum_{\beta}\Lambda_{\alpha \beta} \left( f_{\beta}-f_{\beta}^{eq}
\right)+\sum_{\beta}\left( I_{\alpha
\beta}-\frac{1}{2}\Lambda_{\alpha \beta}\right)S_{\beta} \delta_t .
\label{eq:glbe1}
\end{equation}
Here, the left hand side of Eq.(~\ref{eq:glbe1}) corresponds to the
change in the distribution function during a time interval
$\delta_t$, as particle populations stream from location $\overrightarrow{x}$
to their adjacent location
$\overrightarrow{x}+\overrightarrow{e_{\alpha}}\delta_t$, with a
velocity $\overrightarrow{e_{\alpha}}$ along the characteristic
direction $\alpha$. We consider a three-dimensional, nineteen
velocity (D3Q19) particle velocities set, shown in
Fig.~\ref{fig:d3q19}, given by
%%%%% FIGURE %%%%%
%%%%% FIGURE %%%%%
\begin{equation}
\overrightarrow{e_{\alpha}} = \left\{\begin{array}{ll}
   {(0,0,0)}&{ \alpha=0}\\
   {(\pm 1,0,0),(0,\pm 1,0),(0,0,\pm 1)}&{ \alpha=1,\cdots,6}\\
   {(\pm 1,\pm 1,0),(\pm 1,0,\pm 1),(0,\pm 1,\pm 1)}&{ \alpha=7,\cdots,18.}
\end{array} \right.
\label{eq:velocityd3q19}
\end{equation}
The magnitude of the Cartesian component $c$ of the particle
velocity $\overrightarrow{e_{\alpha}}$ is given by
$c=\delta_x/\delta_t$, where $\delta_t$ is the lattice time step.
%The corresponding column vector of distribution functions
%$\mathbf{f}$ at a location may be written as
%\begin{equation}
%\mathbf{f}=\left[ f_0,f_1,f_2,\ldots,f_{18} \right]^T.
%\label{eq:fvector}
%\end{equation}
%where the superscript 'T' represents the transpose operator.

The first term on the right hand side (RHS) of Eq.~(\ref{eq:glbe1})
represents the cumulative effect of particle collisions on the
evolution of distribution function $f_{\alpha}$. Collision is a
relaxation process in which $f_{\beta}$ relaxes to its local
equilibrium value $f_{\beta}^{eq}$ at a rate determined by the
relaxation time matrix $\Lambda_{\alpha \beta}$. The GLBE has a
generalized collision matrix with multiple relaxation times
corresponding to the underlying physics: the macroscopic fields such
as density, momentum and stress tensors are given as various kinetic
moments of the distribution function. For example, collision does
not alter the densities $\rho$ and momentum $\overrightarrow{j}=\rho
\overrightarrow{u}$, while the stress tensors relax during collisins
at rates determined by fluid properties such as viscosities. The
components of the collision matrix $\Lambda_{\alpha \beta}$ in the
GLBE are developed to reflect the underlying physics of collision as
a relaxation process.

The second term on the RHS of  Eq.(~\ref{eq:glbe1}) introduces
changes in the evolution of distribution function due to external
force fields $\overrightarrow{F}$, such as driving body forces
mimicking a pressure gradient in a periodic domain or gravity, or
Lorentz or Coriolis forces, through a source term $S_{\alpha}$. In this term
$I_{\alpha \beta}$ is the component of the identity matrix $I$. The
source term $S_{\alpha}$ may be written as~\cite{he98,premnath06}
%%%%%%%%%%%%%%%%%%%%%%%% Eqn %%%%%%%%%%%%%%%%%%%%%%%%
\begin{equation}
S_{\alpha}=\frac{(e_{\alpha j}-u_j)F_{j}}{\rho c_s^2}
f_{\alpha}^{eq,M} (\rho,\overrightarrow{u}), \label{eq:sourceterm1}
\end{equation}
%%%%%%%%%%%%%%%%%%%%%%%% Eqn %%%%%%%%%%%%%%%%%%%%%%%%
where $f_{\alpha}^{eq,M}(\rho,\overrightarrow{u})$ is the local
Maxwellian
%%%%%%%%%%%%%%%%%%%%%%%% Eqn %%%%%%%%%%%%%%%%%%%%%%%%
\begin{equation}
f_{\alpha}^{eq,M}(\rho,\overrightarrow{u})= \omega_{\alpha} \rho
\left\{ 1+\frac{\overrightarrow{e_{\alpha}} \cdotp
\overrightarrow{u} }{c_s^2}+
  \frac{\left( \overrightarrow{e_{\alpha}} \cdotp  \overrightarrow{u} \right)^2}{2c_s^4}-
  \frac{1}{2}\frac{ \overrightarrow{u} \cdotp \overrightarrow{u}
  }{c_s^2}\right\}, \omega_{\alpha}=\left\{\begin{array}{ll}
   {\frac{1}{3}}&{ \alpha=0}\\
   {\frac{1}{18}}&{ \alpha=1,\cdots,6}\\
   {\frac{1}{36}}&{ \alpha=7,\cdots,18.}
\end{array} \right.
\label{eq:locMaxwellian}
\end{equation}
%%%%%%%%%%%%%%%%%%%%%%%% Eqn %%%%%%%%%%%%%%%%%%%%%%%%
and $c_s=c/\sqrt{3}$ is the speed of sound of the model. By
neglecting terms of the order of $O(\mathrm{Ma}^2)$ or higher
Eq.(\ref{eq:sourceterm1}) may be simplified as
%%%%%%%%%%%%%%%%%%%%%%%% Eqn %%%%%%%%%%%%%%%%%%%%%%%%
\begin{equation}
S_{\alpha}= w_{\alpha}\left[ \frac{3}{c^2}\left(e_{\alpha
i}-u_i\right)+ \frac{9}{c^4}\left(  \overrightarrow{e_{\alpha}} \cdot  \overrightarrow{u} \right)e_{\alpha i}\right]F_i
\label{eq:simpleforce}
\end{equation}
%%%%%%%%%%%%%%%%%%%%%%%% Eqn %%%%%%%%%%%%%%%%%%%%%%%%
where $F_i=\{F_x, F_y, F_z\}$, with $F_x$, $F_y$ and $F_z$
are the Cartesian components of the external force field, which can, in general, vary in space and/or time.

It may be noted that, Eq.~(\ref{eq:glbe1}) is obtained from the
second-order trapezoidal discretization of the source term in
GLBE~\cite{premnath06}, viz., $
f_{\alpha}(\overrightarrow{x}+\overrightarrow{e_{\alpha}}\delta_t,t+\delta_t)-f_{\alpha}(\overrightarrow{x},t)=
-\sum_{\beta}\Lambda_{\alpha \beta} \left[ f_{\beta}(\overrightarrow{x},t)-f_{\beta}^{eq}(\overrightarrow{x},t)
\right]+1/2[S_{\alpha}(\overrightarrow{x},t) + S_{\alpha}(\overrightarrow{x}+\overrightarrow{e_{\alpha}}\delta_t,t+\delta_t)
]\delta_t$, which is made effectively time-explicit through a
transformation
$\overline{f}_\alpha=f_\alpha-1/2S_\alpha\delta_t$~\cite{he98}, and
then dropping the `overbar' in subsequent representations for
convenience. The second-order discretization provides a more
accurate treatment of source terms, particulary in correctly
recovering general forms of external non-uniform forces in the continuum limit
without spurious terms due to discrete lattice effects~\cite{guo02},
and its time-explicit representation facilitates numerical solution
in a manner analogous to the standard LBE. The local macroscopic
density and velocity fields are given by
%%%%%%%%%%%%%%%%%%%%%%%% Eqn %%%%%%%%%%%%%%%%%%%%%%%%
\begin{equation}
\rho=\sum_{\alpha}f_{\alpha}, \label{eq:density}
\end{equation}
%%%%%%%%%%%%%%%%%%%%%%%% Eqn %%%%%%%%%%%%%%%%%%%%%%%%
%%%%%%%%%%%%%%%%%%%%%%%% Eqn %%%%%%%%%%%%%%%%%%%%%%%%
\begin{equation}
\overrightarrow{j} \equiv
\rho\overrightarrow{u}=\sum_{\alpha}f_{\alpha}\overrightarrow{e_{\alpha}}+\frac{1}{2}\overrightarrow{F}\delta_t,
\label{eq:momentum}
\end{equation}
%%%%%%%%%%%%%%%%%%%%%%%% Eqn %%%%%%%%%%%%%%%%%%%%%%%%
and the pressure field $p$ may be written as
%%%%%%%%%%%%%%%%%%%%%%%% Eqn %%%%%%%%%%%%%%%%%%%%%%%%
\begin{equation}
p=c_s^2\rho. \label{eq:pressure}
\end{equation}
%%%%%%%%%%%%%%%%%%%%%%%% Eqn %%%%%%%%%%%%%%%%%%%%%%%%

The physics behind the kinetic equation Eq.~(\ref{eq:glbe1}), and in
particular, the collision matrix $\Lambda_{\alpha \beta}$ will
become more transparent when it is specified directly in terms of a
set of linearly independent moments $\mathbf{\widehat{f}}$ instead
of the distribution functions $\mathbf{f}=\left[
f_0,f_1,f_2,\ldots,f_{18}\right]^{\dag}$, i.e. through
$\mathbf{\widehat{f}}=\left[
\widehat{f}_0,\widehat{f}_1,\widehat{f}_2,\ldots,\widehat{f}_{18}
\right]^{\dag}$. Here, the superscript `$\dag$' is the transpose
operator and the `hat' represents quantities in moment space. The
moments have direct physical import to the macroscopic quantities
such as momentum and viscous stress tensor. The components of
$\mathbf{\widehat{f}}$ are provided in
Appendix~\ref{app:momentcomponents}. This is achieved through a
transformation matrix $\mathcal{T}$:
$\mathbf{\widehat{f}}=\mathcal{T}\mathbf{f}$. The elements of
$\mathcal{T}$ are given in d'Humi\`{e}res~\emph{et~al.}~\cite{dhumieres02}. Each row of
this matrix is orthogonal to every other row. The essential
principle for its construction is based on the observation that the
collision matrix $\Lambda$ becomes a diagonal matrix
$\widehat{\Lambda}$ through
$\widehat{\Lambda}=\mathcal{T}\Lambda\mathcal{T}^{-1}$ in a suitable
orthogonal basis which can be obtained as combinations of monomials
of the Cartesian components of the particle velocity directions
$\overrightarrow{e_{\alpha}}$ through the standard Gram-Schmidt
procedure.

The collision matrix in moment space $\widehat{\Lambda}$ may thus be
written as
\begin{equation}
\widehat{\Lambda}=diag\left( s_0,s_1,s_2,\ldots,s_{18}\right),
\end{equation}
where $s_0,s_1,s_2,\ldots,s_{18}$ relaxation time rates for the
respective moments. The corresponding components of the local
equilibrium distributions in moment space $
\mathbf{\widehat{f}}^{eq}=\left[
\widehat{f}_0^{eq},\widehat{f}_1^{eq},\widehat{f}_2^{eq},\ldots,\widehat{f}_{18}^{eq}
\right]^{\dag}$ are functions of the conserved moments, viz., local
density and momentum fields, and are given in
Appendix~\ref{app:momentcomponents} .

When there is an external force field, as in
Eq.~(\ref{eq:simpleforce}) represented in particle velocity space
$\mathbf{S}$, where $\mathbf{S}=\left[ S_0,S_1,S_2,\ldots,S_{18} \right]^{\dag}$,
appropriate source terms in moment space
$\mathbf{\widehat{S}}$ need to be introduced. In this regard,
we obtain the projections of source terms onto moments $\mathbf{\widehat{S}}$
by a direct application of the transformation matrix to Eq.~(\ref{eq:simpleforce})
for the D3Q19 model, as done for the D3Q15 model earlier in Ref.~~\cite{premnath06},
i.e. $\mathbf{\widehat{S}}=\mathcal{T}\mathbf{S}$, where $\mathbf{\widehat{S}}=\left[
\widehat{S}_0,\widehat{S}_1,\widehat{S}_2,\ldots,\widehat{S}_{18}
\right]^{\dag}$. They are explicit functions of the
external force field $\overrightarrow{F}$ and the velocity $\overrightarrow{u}$,
which are summarized in Appendix~\ref{app:momentcomponents}.
In effect, due to collisions and the presence of external forces,
the distribution functions in moment space or simply, the moments
are modified by the quantity $-\Lambda\left( \mathbf{\widehat{f}} -
\mathbf{\widehat{f}}^{eq}\right)+\left( \mathrm{I}-1/2\widehat{\Lambda}
\right)\mathbf{\widehat{S}}\delta_t$.

A multiscale analysis based on the Chapman--Enskog
expansion~\cite{chapman64} of the GLBE shows that in the continuum
limit, it corresponds to the weakly compressible Navier--Stokes
equations with external forces, where the density, velocity and
pressure, given by Eqs.~(\ref{eq:density}),~(\ref{eq:momentum}) and
~(\ref{eq:pressure}), respectively, as was done for the D3Q15 model
by Premnath and Abraham~\cite{premnath06}. The macrodynamical equations can also be
derived through an asymptotic analysis under a diffusive
scaling~\cite{sone90,junk01,junk05}. The transport properties of the
fluid flow, such as bulk $\zeta$ and shear $\nu$ kinematic
viscosities can be related to the appropriate relaxation times
through either Chapman--Enskog analysis of the GLBE or the von
Neumann stability analysis of its linearized
version~\cite{lallemand00}:
\begin{eqnarray}
\zeta&=&\frac{2}{9}\left(
\frac{1}{s_1}-\frac{1}{2}\right)\delta_t,\label{eq:bulkvisc}
\\ \nu&=&\frac{1}{3}\left(
\frac{1}{s_{\beta}}-\frac{1}{2}\right)\delta_t, \quad
\beta=9,11,13,14,15.\label{eq:shearvisc}
\end{eqnarray}
Notice that from Eq.~(\ref{eq:shearvisc}),
$s_{9}=s_{11}=s_{13}=s_{14}=s_{15}$ to maintain isotropy of the
stress tensor and $s_2$ determines the magnitude of bulk viscosity.
The rest of the relaxation parameters do not affect hydrodynamics
but can be chosen in such a way to enhance numerical stability as to
simulate higher Reynolds number problems for a given grid
resolution, in particular for wall-bounded turbulent flows considered
here. Based on linear stability analysis~\cite{lallemand00}, the
following values for the other relaxation parameters are
determined~\cite{dhumieres02}: $s_1=1.19$, $s_2=s_{10}=s_{12}=1.4$,
$s_4=s_6=s_8=1.2$ and $s_{16}=s_{17}=s_{18}=1.98$. For the conserved
moments, the values of the relaxation parameters are immaterial as
their corresponding equilibrium distribution is set to the value of
the respective moments itself. However, with forcing terms it is
important that they be non-zero~\cite{mccracken05,premnath06}. For
simplicity, we set $s_0=s_3=s_5=s_7=1.0$. It may be noted that all
relaxation parameters have the following bound $0<s_{\alpha}<2$. In
this paper, we employ the above values for the relaxation
parameters. Since the GLBE employs a set of relaxation times, it is
also referred to as the multiple-relaxation time (MRT)-LBE.

It may be noted that when all the relaxation times are equal, i.e.,
$s_1=s_2=\ldots=s_{18}=1/\tau$, where $\tau$ is the relaxation time,
the GLBE reduces to the single-relaxation time (SRT)-LBE~\cite{qian92,chen92}
based on the Bhatnagar, Gross and Krook model~\cite{bhatnagar54}. Its popularity
and appeal lies in its apparent simplicity. However, the GLBE has marked advantages
when compared with the SRT-LBE: for a given resolution, the GLBE is significantly
more stable numerically and more accurate for problems with anisotropy, with an
insignificant additional computational overhead, thereby allowing
access to a greater range of problems, particularly at higher
Reynolds numbers, to be reached than possible with the SRT-LBE. This
is demonstrated later for two problems involving wall-generated
turbulent flows.

\section{\label{sec:turbmodel} Subgrid Scale Turbulence Model}
In this paper, we have incorporated the subgrid scale (SGS) effects
in the GLBE through the standard Smagorinsky model~\cite{smagorinsky63}.
It assumes that the SGS Reynolds stress term depends on the local strain
rate tensor and leads to the eddy-viscosity assumption. The eddy viscosity
$\nu_t$ arising from this model can be written as
\begin{equation}
\nu_t=\left( C_s \Delta \right)^2\overline{S}, \quad \overline{S}
=\sqrt{2S_{ij}S_{ij}}, \label{eq:sgsmodel}
\end{equation}
where $C_s$ is a constant (taken equal to 0.12 in this paper). Here,
$\Delta$ is the cut-off length scale set equal to the lattice-grid
spacing, i.e. $\Delta=\delta_x$, and $S_{ij}$ is the strain rate
tensor given by $S_{ij}=1/2\left( \partial_j u_i+ \partial_i u_j
\right)$. In LBM, the strain rate tensor can be computed directly
from the non-equilibrium part of the moments, without the need to
apply finite differencing to the velocity field. Recently, Yu \emph{et~al}.~\cite{yu06}
derived such expressions for the strain-rate tensor for the D3Q19 model of the GLBE without forcing term.
In this paper, we extend the results for GLBE with
forcing term by means of a Chapman--Enskog analysis~\cite{chapman64,premnath06}, which is presented in
Appendix~\ref{app:strainrate}. The use of forcing terms allows for incorporation of not only general
forms of non-uniform external forces, but also more general forms of SGS Reynolds stress models~\cite{premnath08a}.
This procedure for calculation of strain rates in GLBE is fully local in space and is computationally
efficient, particularly for complex geometries.

The eddy viscosity $\nu_t$ is added to the molecular viscosity
$\nu_0$, obtained from the statement of the problem, through the
characteristic dimensionless number, such as shear Reynolds number
for turbulent channel flow problem, to yield the total viscosity
$\nu$ (i.e., $\nu=\nu_0+\nu_t$). The relaxation times may then be
obtained from Eq. (\ref{eq:shearvisc}). When such eddy-viscosity
type SGS models are used to provide additional contributions to the
relaxation times, the GLBE can be considered to be ``coarse-grained"
and it can be readily shown that the macroscopic dynamical equations
of fluid flow corresponds to the filtered equations with the SGS
Reynolds stress represented through the eddy viscosity. As a result,
the GLBE would represent the dynamics of larger eddies in turbulent
flows. The distribution functions (or equivalently, the moments) and
the hydrodynamic fields, can be considered to be grid-filtered
quantities. An alternative approach is to directly apply filters to
the moment representation of the GLBE and rigorously derive SGS
models essentially from kinetic theory under appropriate
scaling~\cite{ansumali04}.

To account for the damping of scales near the walls, following an
earlier work~\cite{moin82}, we have implemented the van Driest
damping function~\cite{vandriest56}
\begin{equation}
\Delta=\delta_x\left[1-\exp\left( -\frac{z^{+}}{A^{+}}\right)
\label{eq:vanDriestdamping} \right]
\end{equation}
where $z^{+}$ is the distance from the wall and $A^{+}$ is a
constant, taken to be $25$ in this work~\cite{moin82}. The
superscript $+$ signifies normalization with respect to wall units,
i.e. $z^{+}=z/\delta_{\nu}$, where $\delta_{\nu}=u_{*}/\nu_0$ is the
characteristic viscous length scale. Here, $u_*$ is the shear or friction velocity, which
is related to the wall shear stress $\tau_w$ through $u_*=\sqrt{\tau_w/\rho}$.
While this approach has some empiricism built-in, for a class of
wall-bounded turbulent flows, such as turbulent channel flows
considered here, it has been shown to be reasonably accurate
in prior work based on the solution of grid-filtered
Navier--Stokes equations~\cite{moin82}. Also, as will be shown later
in this paper that the GLBE is able to reproduce turbulence statistics
in the near-wall region reasonable well using this damping supplemented
to the SGS model. For more general situations, for improved accuracy
it may be necessary to introduce dynamic SGS models
(e.g., ~\cite{germano91,zhang93,salvetti95}) for LES
using the GLBE~\cite{premnath08a}.

\section{\label{sec:compoptimize} Computational Procedure: Optimization with Forcing Terms}
In practice, implementation of the GLBE with forcing term, i.e.
Eq.~(\ref{eq:glbe1}), together with associated turbulence models and
procedure for strain rate computations, initial and boundary
conditions, requires careful consideration for the details for
efficient performance. In particular, the ``effective" collision step including the
forcing terms should be performed in \emph{moment space}, while the streaming step should be
executed in \emph{particle velocity space} and the special
properties of the transformation matrix that transform between the
two spaces should be fully
exploited~\cite{dhumieres02,yu06,premnath06}. Such properties of the
transformation matrix $\mathcal{T}$ include its orthogonality,
entries with many zero elements, and entries with many common
elements that are integers, which are used to form the most compact
common sub-expressions for transformations between spaces. We will now
briefly discuss the details of the computational procedure for the GLBE with
forcing term used in this paper.

The GLBE with forcing term can be re-written in terms of the
following ``effective" collision and streaming steps, respectively:
\begin{equation}
\mathbf{\widetilde{f}}(\overrightarrow{x},t)=\mathbf{f}(\overrightarrow{x},t)+\boldsymbol{\varpi}(\overrightarrow{x},t),
\label{eq:postcollision}
\end{equation}
and
\begin{equation}
f_{\alpha}(\overrightarrow{x}+\overrightarrow{e}_{\alpha}\delta_t,t+\delta_t)=f_{\alpha}(\overrightarrow{x},t),
\label{eq:streaming}
\end{equation}
where $\widetilde{f}_{\alpha}$ is the post-collision distribution
function and
\begin{equation}
\boldsymbol{\varpi}(\overrightarrow{x},t)= \mathcal{T}^{-1}\left[
-\widehat{\Lambda}\left(\mathbf{\widehat{f}}-\mathbf{\widehat{f}}^{eq}
\right)+\left(\mathcal{I}-\frac{1}{2}\widehat{\Lambda}
\right)\mathbf{\widehat{S}} \right],
\label{eq:relax_term}
\end{equation} is the effective change due to collision including the effect
of external forces. Here, $\mathbf{\widehat{f}}\equiv\mathbf{\widehat{f}}(\overrightarrow{x},t)$,
$\mathbf{\widehat{f}}^{eq}\equiv\mathbf{\widehat{f}}^{eq}(\overrightarrow{x},t)$ and
$\mathbf{\widehat{S}}\equiv\mathbf{\widehat{S}}(\overrightarrow{x},t)$, and $\mathcal{I}$ is the
identity matrix and $\widehat{\Lambda}=\mathcal{T}\Lambda\mathcal{T}^{-1}=diag(s_0,s_1,\ldots,s_{18})$
is the diagonal collision matrix in moment space.

A note regarding the actual implementation details is in order. First, the transformation matrix
$\mathcal{T}$ is row-wise orthogonal and satisfies $\mathcal{T}\mathcal{T}^{\dag}=\mathcal{\widehat{L}}$,
where $\mathcal{T}^{\dag}$ is the transpose of $\mathcal{T}$ and $\mathcal{\widehat{L}}$ is a diagonal
normalization matrix. Thus it follows that the matrix inverse $\mathcal{T}^{-1}$ is obtained simply using $\mathcal{T}^{-1}=\mathcal{T}^{\dag}\mathcal{\widehat{L}}$. As a result, we may write Eq.~(\ref{eq:relax_term})
as $\boldsymbol{\varpi}(\overrightarrow{x},t)=\mathcal{T}^{\dag}\mathbf{\widehat{q}}$
where $\mathbf{\widehat{q}}$ is given by
$\mathbf{\widehat{q}}= \left[
-\widehat{\Gamma}\left(\mathbf{\widehat{f}}-\mathbf{\widehat{f}}^{eq}
\right)+\left(\mathcal{\widehat{L}}^{-1}-\frac{1}{2}\widehat{\Gamma}\right)\mathbf{\widehat{S}}\right]$ and $\widehat{\Gamma}= \mathcal{\widehat{L}}^{-1}\widehat{\Lambda}$. Thus, for computational efficiency, we actually
implement the ``effective" collision step that also including forcing terms in moment space.
Now, the relaxation times in $\widehat{\Lambda}$ used to compute in
Eq.~(\ref{eq:relax_term}) can be related to the transport
coefficients and modulated by eddy viscosity, in the case of
hydrodynamic time scales, as follows: $ s_{1}^{-1}=
s_{\zeta}^{-1}=\frac{9}{2}\zeta+\frac{1}{2}$ from
Eq.~(\ref{eq:bulkvisc}), and
$s_{9}=s_{11}=s_{13}=s_{14}=s_{15}=s_{\nu}$, where $ s_{\nu}^{-1}=
3\nu+\frac{1}{2}= 3(\nu_0+\nu_t)+\frac{1}{2}$, from
Eq.~(\ref{eq:shearvisc}). The eddy viscosity $\nu_t$ is obtained
from Eq.~(\ref{eq:sgsmodel}). The rest of the relaxation parameters
can be chosen to enhance numerical stability, as discussed in
Section~\ref{sec:glbe}. The forcing term used in the computation of strain rate tensor (Appendix~\ref{app:strainrate})
and in the ``effective" collision step (Eq.~(\ref{eq:relax_term})) can be obtained from
Appendix~\ref{app:momentcomponents}. This optimized procedure dramatically improves the computational speed of the
GLBE as compared to a naive implementation. Indeed, the additional computational overhead of using GLBE in lieu
of the SRT-LBE is small, between $15\%-30\%$, but, as will be shown later, with a significantly
improved accuracy and numerical stability.

No slip wall boundary conditions, involving stationary walls as well
as moving walls, in the case of turbulent channel flow and driven
cavity flow, respectively, are implemented by means of the link or
half-way bounce back~\cite{ladd94}. To initiate turbulence, a three-dimensional perturbation velocity field
satisfying divergence free condition~\cite{lam89} is employed in the solution of the GLBE through the consistent
initialization procedure~\cite{mei06}.

\section{\label{sec:turbchannelflow}Fully-Developed Turbulent Channel Flow}
First, we simulated a canonical problem, viz., fully-developed
turbulent channel flow using the GLBE with the SGS model mentioned
above. Prior efforts have validated LBM as a DNS tool for this
problem by comparing a set of turbulent statistics with available
data~\cite{amati97,amati99,lammers06}. The focus of this study is to
evaluate MRT-LBE that incorporate subgrid scale effects for this
problem on a relatively coarse grid, while maintaining the necessary
near-wall resolution.

\begin{figure}
\includegraphics[width = 120mm,viewport=50 130 720
500,clip]{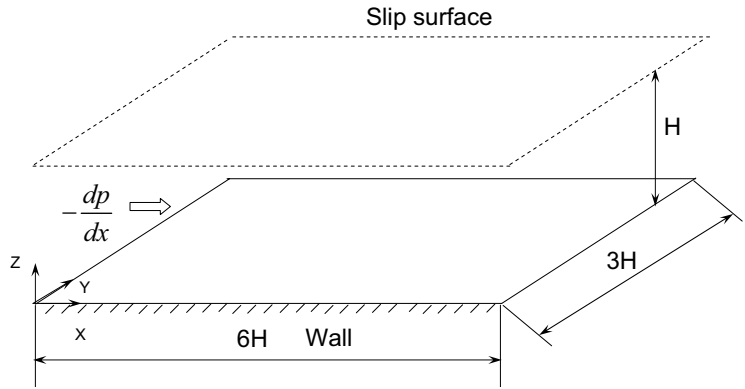}% Here is how to import EPS art
\caption{\label{fig:channelschematic} Schematic of computational
domain for LES of fully-developed turbulent channel flow.}
\end{figure}

We considered turbulent flow with a shear
Reynolds number $\mathrm{Re}_{*}=u_{*}H/\nu_{0}=183.6$, where $\nu_0$ is the
molecular kinematic viscosity and $H$ is the channel half height. A
schematic of the problem set up is shown in
Fig.~\ref{fig:channelschematic}, in which a no-slip boundary is
imposed at the bottom, free-slip at the top and periodic boundary
conditions were applied in the streamwise $x$ and spanwise $y$ directions~\cite{lam89,
lam92}.
%%%%% FIGURE %%%%%
%%%%% FIGURE %%%%%
The computational domain is chosen with appropriate aspect ratios,
viz., $6H$ and $3H$ in the streamwise and spanwise directions,
respectively. With this domain, a sufficient number of wall-layer
streaks are accommodated~\cite{lam89} and end effects of two-point correlations
are excluded, i.e. the two-point velocity correlations in solutions
are required to decay nearly to zero within half the
domain~\cite{moin98}. For this initial case, we considered a uniform
grid with a grid spacing in wall units (referred to with a ``$+$"
superscript) as $\Delta^{+} = \Delta/\delta_{\nu}=4.08$, where
$\delta_{\nu}=\nu_{0}/u_{*}$ is the viscous length
scale as defined in Sec.~\ref{sec:turbmodel}.
The computational domain thus consists of $270\times
135\times 47$ grid nodes. Due to the use of link-bounce back method
for implementation of wall boundary condition, the first lattice
node is located at a distance of $\Delta^{+}_{nw}=\Delta^{+}/2$,
which in our case is 2.04. For wall-bounded turbulent flows, it is
important to adequately resolve the near-wall, small-scale turbulent
structures, which is satisfied when the computations resolve the
local dissipative or Kolmogorov length scale
$\eta=(\nu^3_{0}/\epsilon)^{1/4}$, i.e. $\Delta^{+}_{nw}\leq
O(\eta^{+})$~\cite{moin98}. In particular, it is generally
recognized that $1.5\eta^{+}-2.0\eta^{+}$ represents the upper limit
of grid-spacing, above which the small scale turbulent motions in
bounded flows are not well resolved. It can be shown by simple
arguments that $\eta^{+}\approx 1.5-2.0$ at the wall and that
$\eta^{+}$ increases with increasing distance from the
wall~\cite{pope00}. Thus, our computational set up is expected to
fairly resolve the small-scale turbulent structures.

The initial mean velocity is specified to satisfy the $1/7^{th}$
power law~\cite{pope00}, while initial perturbations satisfying
divergence free velocity field~\cite{lam89}. The density field is
taken to be $\rho=\rho_0=1.0$ for the entire domain. The precise
form of the initial fields may not affect affect the turbulence
statistics, but can have significant influence on the number of time steps
needed for convergence of
the solution to statistically steady state. In particular, the above
choice of initial fields would enable a rapid convergence to the
statistically steady state solution of the fluctuating fields
obtained by the GLBE with forcing term. With these initial conditions on
the macroscopic fields, we employed the consistent initialization
procedure for the distribution functions or moments~\cite{mei06}. Using
$\overrightarrow{F}=-\frac{dp}{dx}\widehat{x}=\frac{\tau_w}{H}\widehat{x}=\frac{\rho
u_{*}^2}{H}\widehat{x}$ as the driving force, the GLBE computations
are carried out until stationary turbulence statistics are obtained,
as measured by the invariant Reynolds stresses profiles. This
initial run was carried out for a duration of $50T^{*}$, where
$T^{*}=H/u_{*}$ is the characteristic time scale. The averaging of
various flow quantities was carried out in time as well as in space
in the homogeneous directions, i.e. over the horizontal planes, by
an additional run for a period of $30T^{*}$.

\begin{figure}
\includegraphics[width = 100mm,viewport=0 0 730
525,clip]{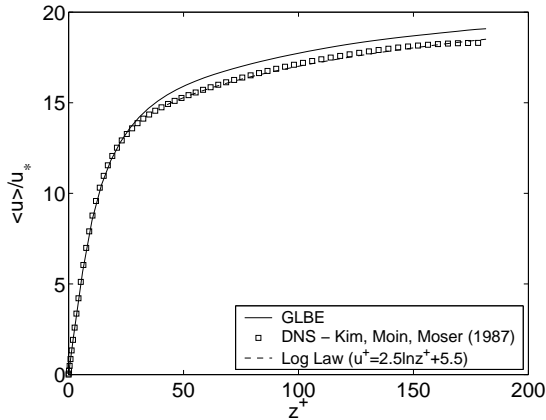}% Here is how to import EPS art
\caption{\label{fig:meanvel} Comparison of the computed mean
velocity profile with wall-layer scaling laws in outer wall
coordinates for fully-developed turbulent channel flow at $\mathrm{Re}_{*} =
183.6$.}
\end{figure}

Figure~\ref{fig:meanvel} shows the computed mean velocity profile,
normalized by the shear velocity $u_{*}$, as a function of the
distance from the wall given in wall units, i.e.
$z^{+}=z/\delta_{\nu}$, where $\delta_{\nu}$ is the viscous length
scale defined earlier.
%\begin{figure}
%\includegraphics[width = 100mm,viewport=80 120 730
%525,clip]{MeanVel}% Here is how to import EPS art
%\caption{\label{fig:meanvel} Comparison of the computed mean
%velocity profile with wall-layer scaling laws in outer wall
%coordinates for fully-developed turbulent channel flow at $Re_{*} =
%183.6$.}
%\end{figure}
%%% FIGURE %%%%%
%%%%% FIGURE %%%%%
Also plotted are the DNS data by Kim, Moin and Moser (1987)~\cite{kim87} based on
a spectral method and the von Karman log-law of the wall,
which is valid for the so-called log-region. The computed velocity
profile follows the DNS data fairly closely, with about $5\%$
difference. Such differences are characteristic of LES, which employ
relatively coarser grids than DNS, and they also generally depend
on the numerical dissipation of the computational approach for LES
(see e.g., Ref.~\cite{choi00,gullbrand03}).

\begin{figure}
\includegraphics[width = 110mm,viewport=0 0 700
550,clip]{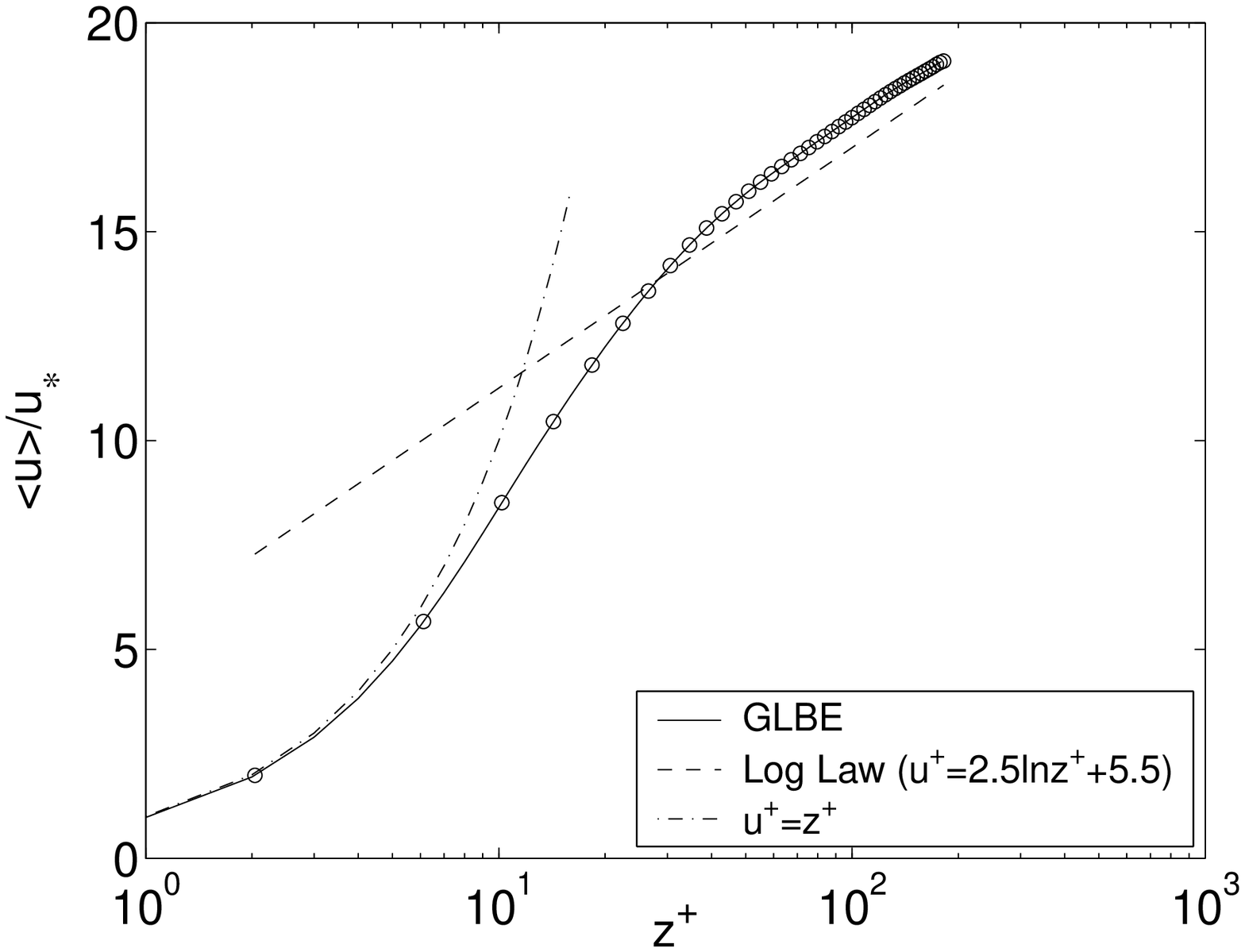}% Here is how to import EPS art
\caption{\label{fig:meanvellog} Comparison of the computed mean
velocity profile with wall-layer scaling laws in inner wall
coordinates for fully-developed turbulent channel flow at $\mathrm{Re}_{*} =
183.6$.}
\end{figure}

Figure~\ref{fig:meanvellog} shows comparison of the computed mean
velocity profile as a function of the distance from the wall with
wall-layer scaling laws, i.e. viscous sublayer and log-law of the
wall.
%\begin{figure}
%\includegraphics[width = 110mm,viewport=20 70 700
%550,clip]{MeanVelLog}% Here is how to import EPS art
%\caption{\label{fig:meanvellog} Comparison of the computed mean
%velocity profile with wall-layer scaling laws in inner wall
%coordinates for fully-developed turbulent channel flow at $Re_{*} =
%183.6$.}
%\end{figure}
%%%%%% FIGURE %%%%%
%%%%% FIGURE %%%%%
Generally, Reynolds stress effects are negligible in the viscous
sublayer region ($z^{+}\leq 5$), and $u^{+}\equiv <u>/u_{*}=z^{+}$
holds for the mean velocity. For $z^{+}>30$, the mean velocity
satisfies the log-law, i.e. $u^{+}=A\mathrm{ln}z^{+}+B$, where the coefficients
depend on the flow parameters and nature of the wall. Values of
$A=2.5$ and $B=5.5$ are known to be reasonably accurate for flow
over smooth walls at $\mathrm{Re}_{*} \approx 180$~\cite{lam89,kim87,pope00}. It can
be seen that computations agree well with these scaling laws.

\begin{figure}
\includegraphics[width = 90mm,viewport=0 0 540
720,clip]{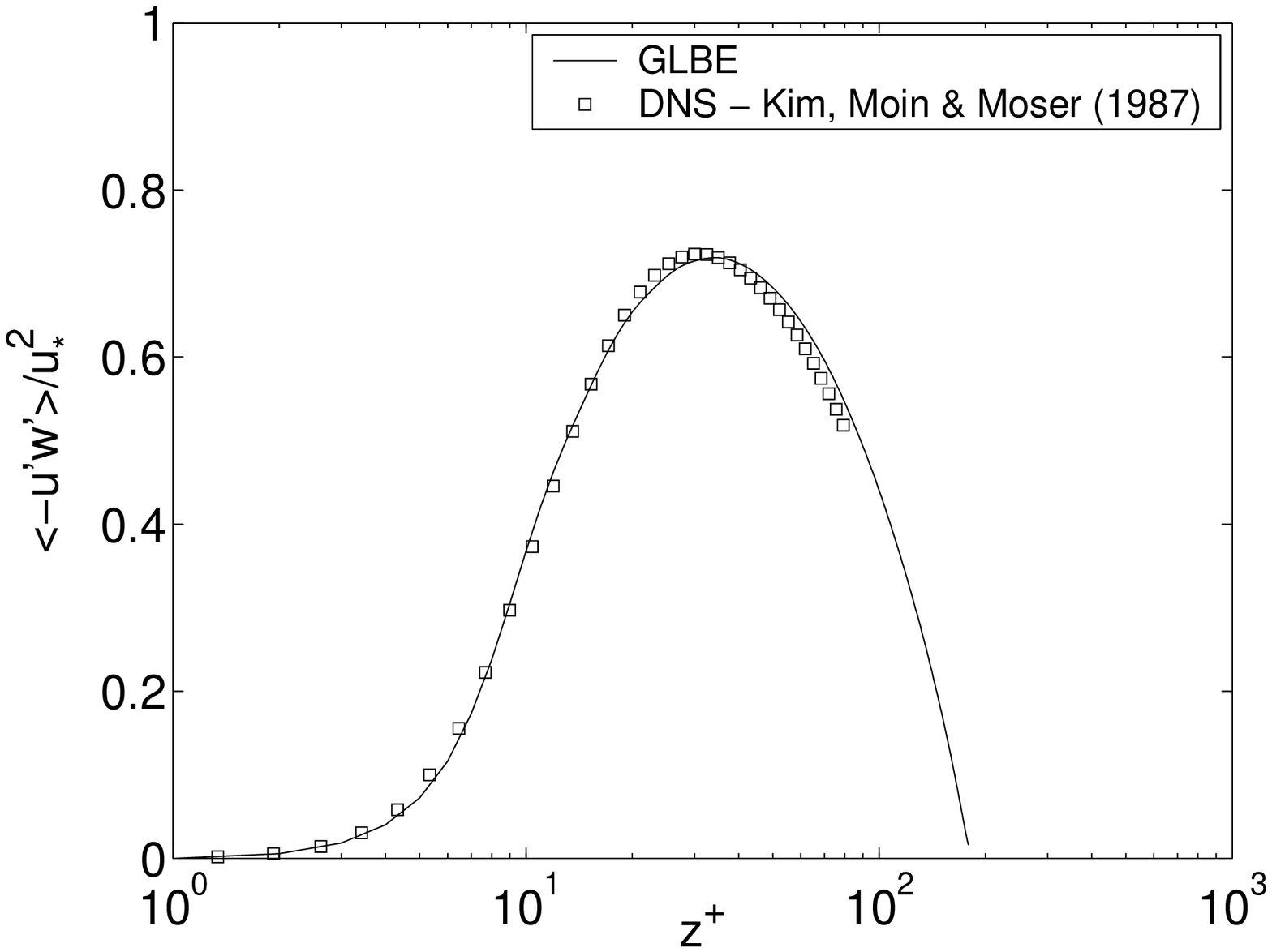}% Here is how to import EPS art
\caption{\label{fig:reynoldsstress} Reynolds stress normalized by
the wall shear stress for fully-developed turbulent channel flow at
$\mathrm{Re}_{*} = 183.6$.}
\end{figure}

The Reynolds stress, normalized by the wall-shear stress, is
presented in Fig.~\ref{fig:reynoldsstress} in semi-log scale and
compared with the DNS data of Kim, Moin and Moser~\cite{kim87} obtained from the
direct solution of incompressible Navier--Stokes equations (NSE),
which the GLBE computations reproduce quite well with good accuracy.
%\begin{figure}
%\includegraphics[width = 120mm,viewport=80 435 540
%720,clip]{Reynoldsstress}% Here is how to import EPS art
%\caption{\label{fig:reynoldsstress} Reynolds stress normalized by
%the wall shear stress for fully-developed turbulent channel flow at
%$Re_{*} = 183.6$.}
%\end{figure}
%%%% FIGURE %%%%%
%%%%% FIGURE %%%%%

\begin{figure}
\includegraphics[width = 120mm,viewport=0 0 700
530,clip]{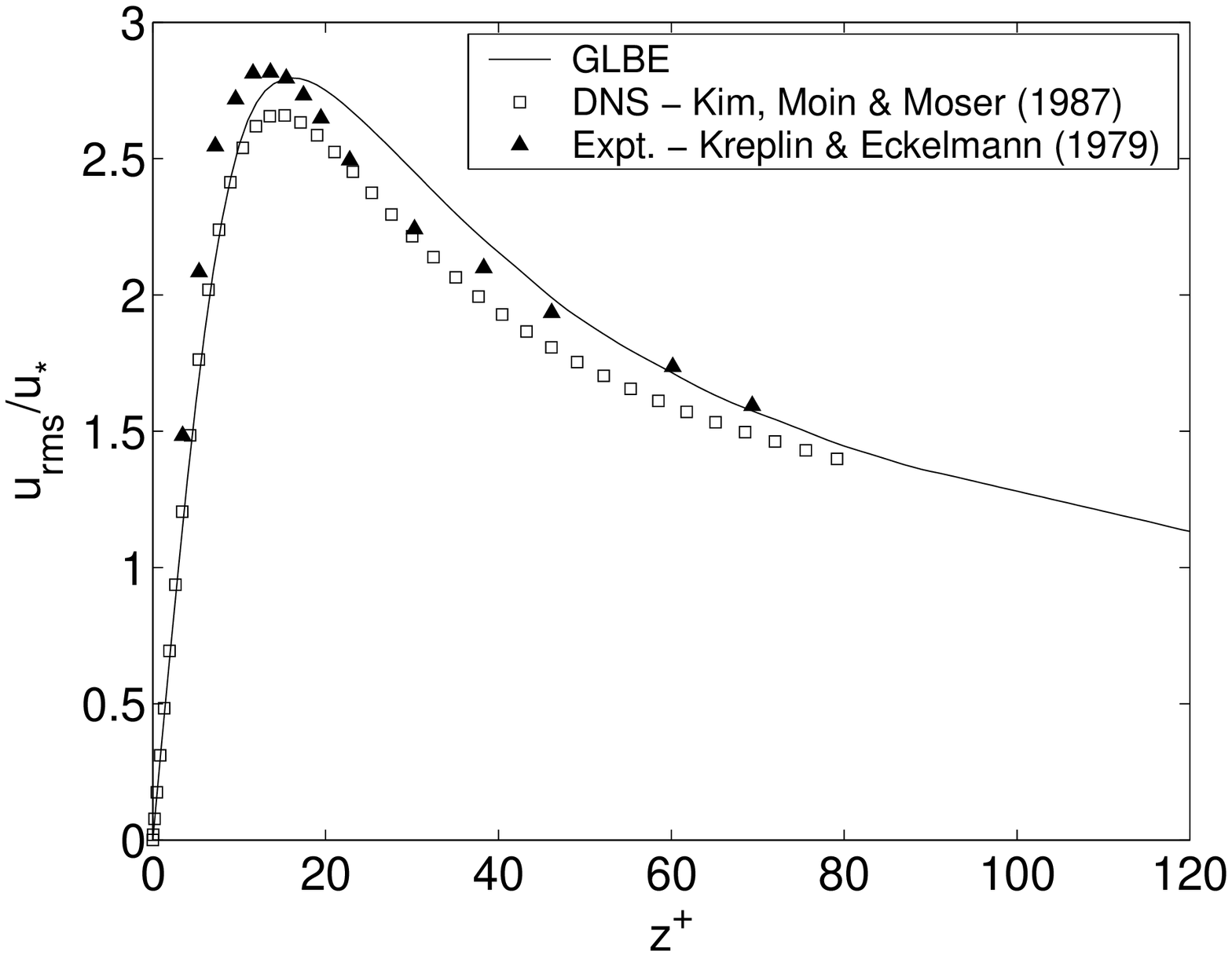}% Here is how to import EPS art
\caption{\label{fig:urms} Root-mean-square (rms) streamwise velocity
fluctuations normalized by the wall shear velocity for
fully-developed turbulent channel flow at $\mathrm{Re}_{*} = 183.6$.}
\end{figure}

\begin{figure}
\includegraphics[width = 120mm,viewport=0 0 700
530,clip]{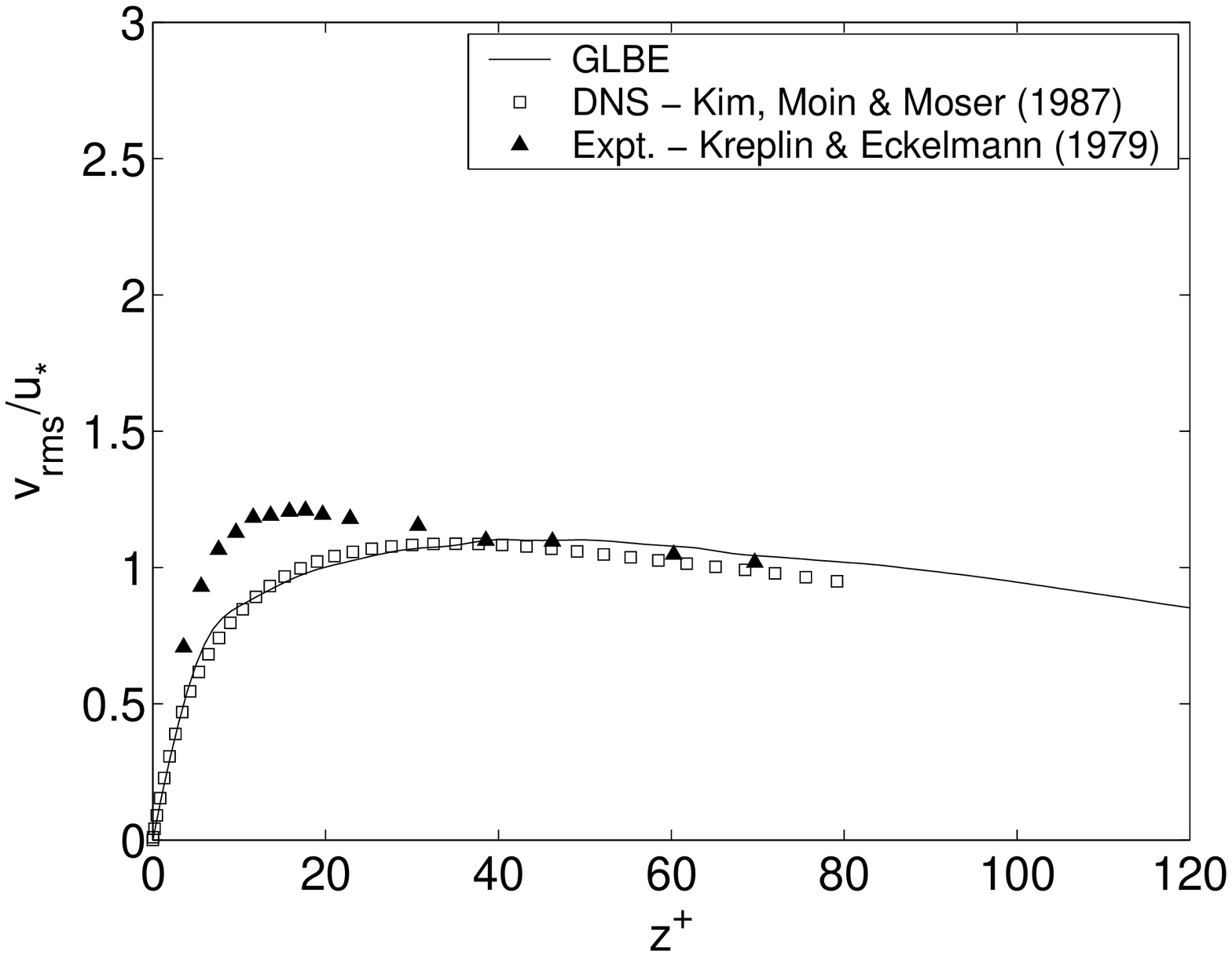}% Here is how to import EPS art
\caption{\label{fig:vrms} Root-mean-square (rms) spanwise velocity
fluctuations normalized by the wall shear velocity for
fully-developed turbulent channel flow at $\mathrm{Re}_{*} = 183.6$.}
\end{figure}

\begin{figure}
\includegraphics[width = 120mm,viewport=0 0 700
530,clip]{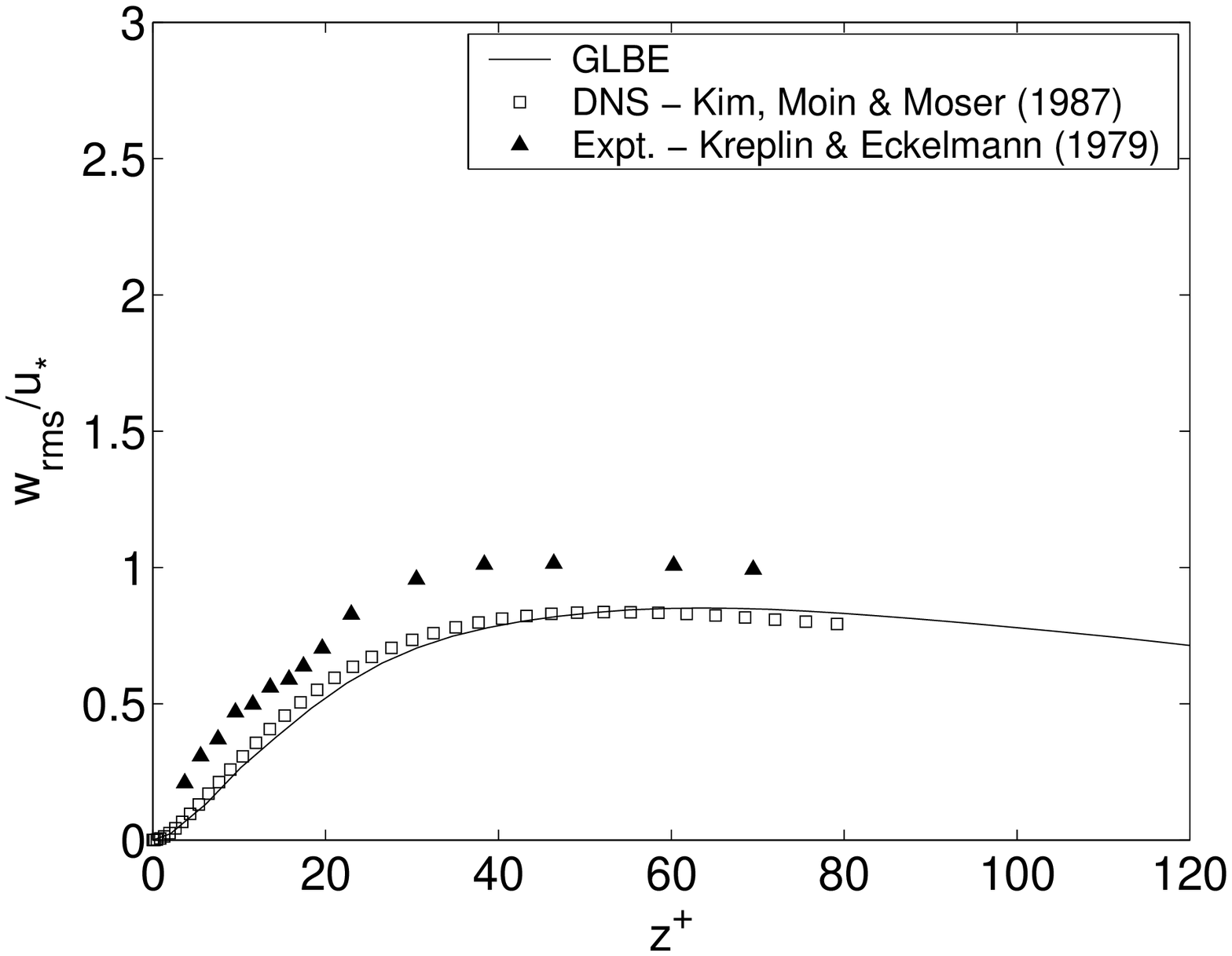}% Here is how to import EPS art
\caption{\label{fig:wrms} Root-mean-square (rms) wall-normal
velocity fluctuations normalized by the wall shear velocity for
fully-developed turbulent channel flow at $\mathrm{Re}_{*} = 183.6$.}
\end{figure}

Let us now consider the statistics of turbulent fluctuations of
important quantities in the near-wall region.
Figures~\ref{fig:urms}, \ref{fig:vrms} and \ref{fig:wrms} show
comparisons of the components of the root-mean-square (rms)
streamwise, spanwise and wall-normal velocity fluctuations,
respectively, computed using the GLBE with data from DNS based on
the solution of NSE by Kim, Moin and Moser~\cite{kim87} and
experimental measurements of Kreplin and Eckelmann~\cite{kreplin79}. It
may be seen that the computed results agree reasonably well with
prior data.
%\begin{figure}
%\includegraphics[width = 120mm,viewport=60 100 700
%530,clip]{urms}% Here is how to import EPS art
%\caption{\label{fig:urms} Root-mean-square (rms) streamwise velocity
%fluctuations normalized by the wall shear velocity for
%fully-developed turbulent channel flow at $Re_{*} = 183.6$.}
%\end{figure}
%%%%% FIGURE %%%%%
%%%%%% FIGURE %%%%%
%\begin{figure}
%\includegraphics[width = 120mm,viewport=15 45 730
%560,clip]{vrms}% Here is how to import EPS art
%\caption{\label{fig:vrms} Root-mean-square (rms) spanwise velocity
%fluctuations normalized by the wall shear velocity for
%fully-developed turbulent channel flow at $Re_{*} = 183.6$.}
%\end{figure}
%%%%%% FIGURE %%%%%
%% FIGURE %%%%%
%%%% FIGURE %%%%%
%\begin{figure}
%\includegraphics[width = 120mm,viewport=20 55 730
%560,clip]{wrms}% Here is how to import EPS art
%\caption{\label{fig:wrms} Root-mean-square (rms) wall-normal
%velocity fluctuations normalized by the wall shear velocity for
%fully-developed turbulent channel flow at $Re_{*} = 183.6$.}
%\end{figure}
%%%%%% FIGURE %%%%%
%%%% FIGURE %%%%%

\begin{figure}
\includegraphics[width = 120mm,viewport=0 0 700
545,clip]{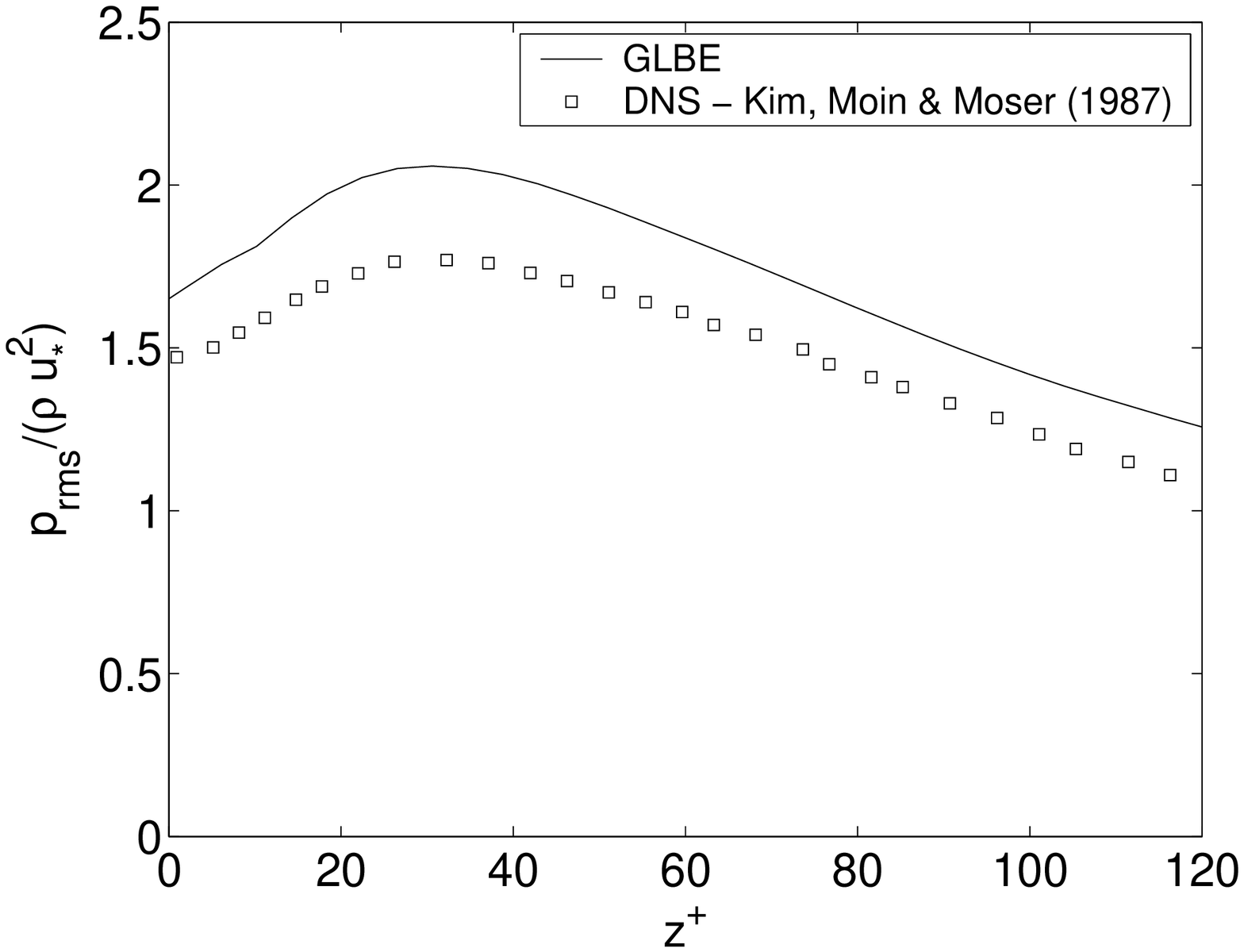}% Here is how to import EPS art
\caption{\label{fig:prms} Root-mean-square (rms) pressure
fluctuations normalized by the wall shear stress for fully-developed
turbulent channel flow at  $\mathrm{Re}_{*} = 183.6$.}
\end{figure}

Another important quantity representing turbulent activity near the
wall is the pressure fluctuations. Figure~\ref{fig:prms} shows the
computed rms pressure fluctuations. The profile shown here is
qualitatively consistent with the NSE-DNS results.
%\begin{figure}
%\includegraphics[width = 120mm,viewport=70 90 700
%545,clip]{prms}% Here is how to import EPS art
%\caption{\label{fig:prms} Root-mean-square (rms) pressure
%fluctuations normalized by the wall shear stress for fully-developed
%turbulent channel flow at  $Re_{*} = 183.6$.}
%\end{figure}
%%%%%% FIGURE %%%%%
%%%%% FIGURE %%%%%
It is found that the pressure fluctuations normalized by the wall
shear stress is about 1.66 at the wall, which is within the range in
prior data -- NSE based DNS results in Ref.~\cite{kim87}
and~\cite{lam89} provide values of about 1.5 and 2.15 respectively.
These values depend on the Reynolds number employed. In the
measurements reported by Willmarth~\cite{willmarth75}, the values of
maximum rms pressure fluctuations were found to be between 2 and 3,
but these were for much higher Reynolds numbers than considered
here. Moreover, the computed maximum pressure fluctuations occurs at
$z^{+} \approx 26$, which is close to the range in DNS
data~\cite{kim87}, i.e. $z^{+} \approx 30$. It may be seen that
there the computed profile rms pressure fluctuations using GLBE is
systematically somewhat larger than the DNS results based on NSE.
This observation is also consistent with those found in
Ref.~\cite{lammers06}, where DNS using SRT-LBE revealed similar
values for the peak rms pressure fluctuations and its location. Such
difference could plausibly be due to compressibility effects
inherent in LBM, while the DNS carried out in Ref.~\cite{kim87}
considered incompressible NSE.

\begin{figure}
\includegraphics[width = 120mm,viewport=0 0 700
500,clip]{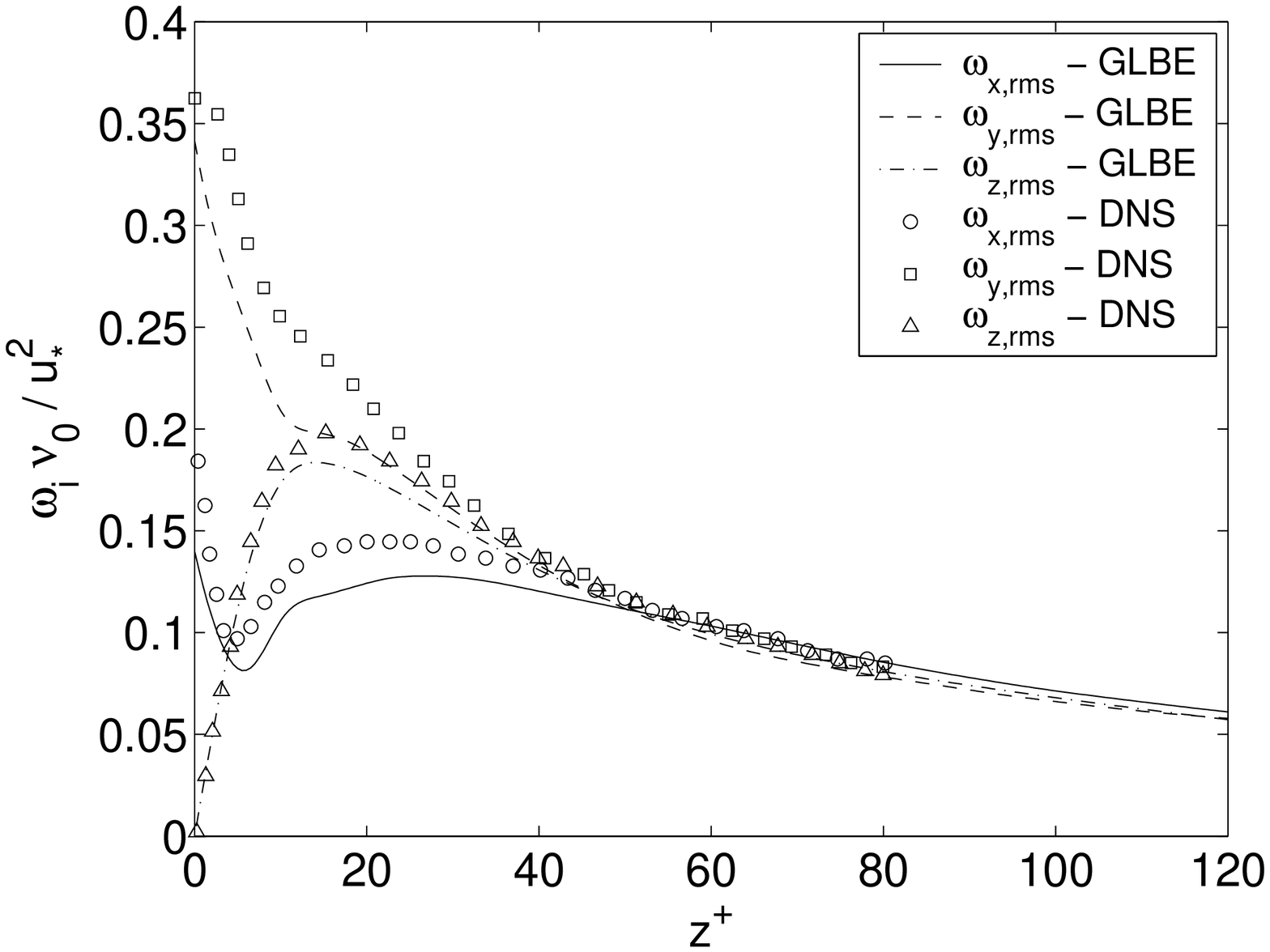}% Here is how to import EPS art
\caption{\label{fig:vortrms} Components of root-mean-square (rms)
vorticity fluctuations normalized by a factor given in terms of wall
shear stress for fully-developed turbulent channel flow at $\mathrm{Re}_{*} =
183.6$. Lines are GLBE solution and symbols are DNS
data~\cite{kim87}.}
\end{figure}

A particularly stringent test is the comparison of computed
components of near wall rms vorticity fluctuations with DNS, which
is shown in Fig.~\ref{fig:vortrms}, in which lines represent the GLBE
solution and symbols the DNS data~\cite{kim87}. The components of
vorticity fluctuations normalized by the mean wall shear
($u_*^2/\nu_0$).
%\begin{figure}
%\includegraphics[width = 120mm,viewport=70 90 720
%500,clip]{vortrms}% Here is how to import EPS art
%\caption{\label{fig:vortrms} Components of root-mean-square (rms)
%vorticity fluctuations normalized by a factor given in terms of wall
%shear stress for fully-developed turbulent channel flow at $Re_{*} =
%183.6$. Lines are GLBE solution and symbols are DNS
%data~\cite{kim87}.}
%\end{figure}
%%%%% FIGURE %%%%%
There is a strong variation among the components of vorticity due to
inhomogeneity and anisotropy of turbulence closer to the wall. Also,
as expected, for distances further from the wall, all the components
of vorticity are essentially the same. While there are some
deviations from the DNS results, the GLBE is able to reproduce the
qualitative trends and, more importantly, the ratio between
components of vorticity. Such deviations have been observed in LES
using filtered NSE, e.g.~\cite{dubief00}, who point out the
underestimation of the resolved components to be due to mere
consequence of filtering process inherent to LES.

\begin{figure}
\includegraphics[width = 120mm,viewport=0 0 700
500,clip]{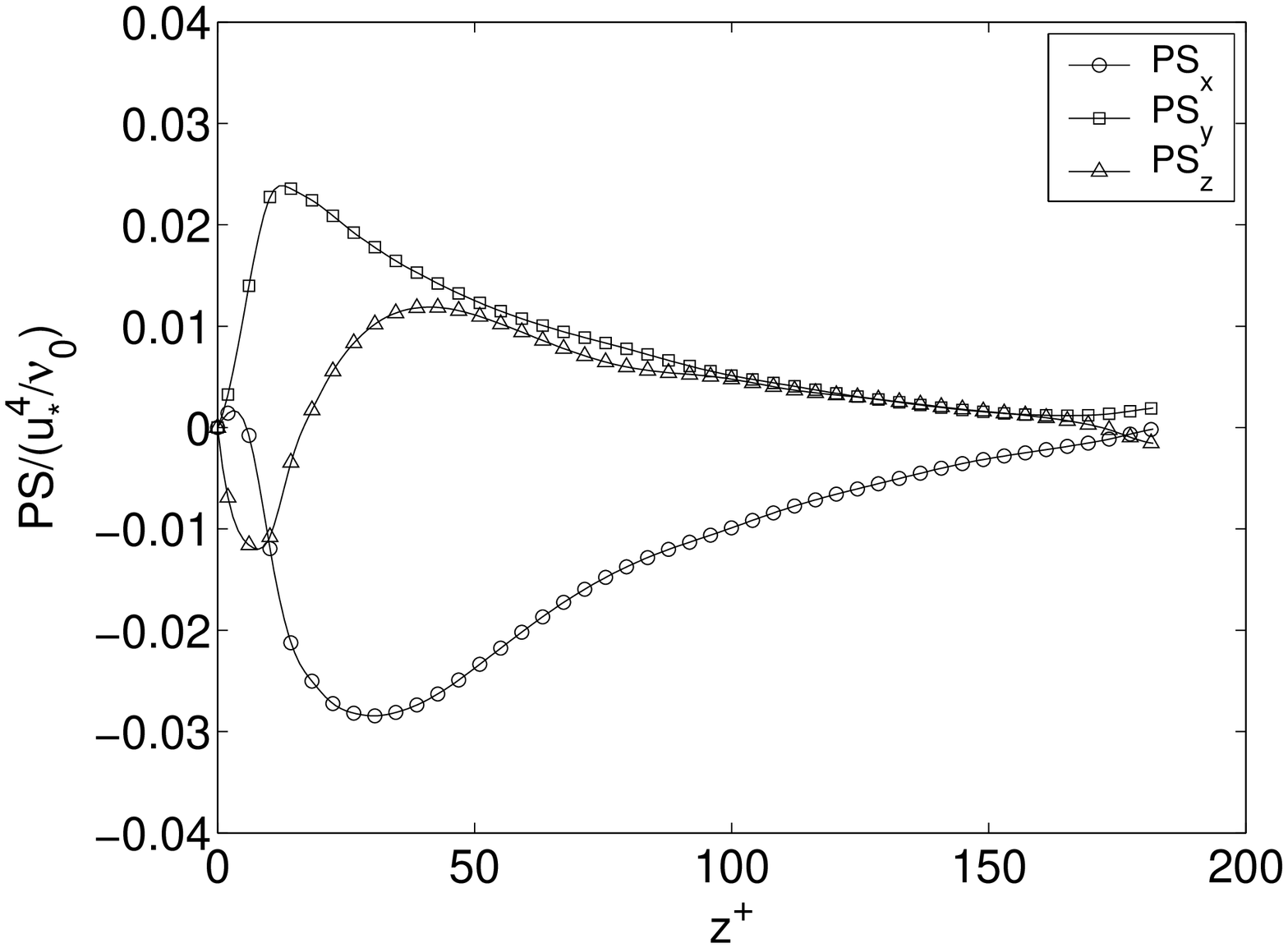}% Here is how to import EPS art
\caption{\label{fig:presstrain} Components of computed
pressure-strain correlations normalized by a factor given in terms
of wall shear velocity and molecular viscosity for fully-developed
turbulent channel flow at $\mathrm{Re}_{*} = 183.6$.}
\end{figure}

Another important measure is the pressure-strain (PS) correlations.
Their components are: $PS_x=\langle p^{'}\partial_x u_x^{'}\rangle$,
$PS_y=\langle p^{'}\partial_y u_y^{'}\rangle$, and $PS_z=\langle
p^{'}\partial_z u_z^{'}\rangle$, where the prime denotes
fluctuations and the brackets refer to averaging (along homogeneous
spatial directions and time). They provide indications of energy
transfers among the components. Figure~\ref{fig:presstrain} shows
the components of the computed PS correlations.
%\begin{figure}
%\includegraphics[width = 120mm,viewport=60 100 700
%500,clip]{presstrain}% Here is how to import EPS art
%\caption{\label{fig:presstrain} Components of computed
%pressure-strain correlations normalized by a factor given in terms
%of wall shear velocity and molecular viscosity for fully-developed
%turbulent channel flow at $Re_{*} = 183.6$.}
%\end{figure}
%%%%%% FIGURE %%%%%
%%%%% FIGURE %%%%%
They exhibit the expected behavior close to the wall, including the
transfer of energy from the wall-normal component to the other two
components near the wall -- a phenomenon termed as splatting or
impingement~\cite{moin82}. Thus, it appears that the GLBE with forcing
term is a reliable approach for computation of fully-developed
turbulent channel flows.

\subsection{\label{sec:stabilityturbchannelflow}Numerical Stability}
To put things in perspective, let us now discuss the stability
characteristics of the GLBE in relation to the SRT-LBE for turbulent
channel flow on coarser grids. The test case used a shear Reynolds
number $\mathrm{Re}_{*}$ of $180$, with a uniform spacing $\Delta^{+}$ of $6$
in wall units, which at the near-wall node becomes
$\Delta^{+}_{nw}=3$ due to the link-bounce back scheme employed. The
number of grid nodes used in each case is $180 \times 90 \times 32$.
This is a somewhat coarser resolution than used in the previous
simulation and it is expected that small-scale near-wall dynamics may not be
properly resolved. Nevertheless, subgrid scale motions are quite
energetic for such coarse resolutions and it is important to
determine if the grid scale numerical instabilities developed by the
computational approaches interact with them. The numerical stability
of the LBM depends on various factors including the grid resolution
$\Delta$, maximum velocity or Mach number $\mathrm{Ma}$ considered and the
relaxation times or the molecular viscosity of the fluid $\nu_0$.
For a given resolution and maximum flow velocity, the numerical
stability of the LBE depends mainly on the molecular viscosity of the
fluid $\nu_0$.

As is natural for LBE, unless otherwise specified, all the results
are reported in lattice units. That is, the velocities are scaled by
the particle velocity $c$ and the distance by the streaming distance
of the populations, $\delta_x$. Here, we considered a maximum
velocity, i.e. velocity at the top surface to be about $0.18$, and
varied the viscosity $\nu_0$. In the case of the SRT-LBE, the only
parameter that can be used to specify $\nu_0$ is the
single-relaxation time $\tau$ and its value is chosen from
$\nu_0=1/3(\tau-1/2)$. On the other hand, for the GLBE, the
relaxation parameters that determine moments involving fluid
stresses are determined from Eq.~\ref{eq:shearvisc}, while the rest
of the parameters are tuned to improve numerical stability as
specified earlier.

\begin{figure}
\includegraphics[width = 120mm,viewport=0 0 700
450,clip]{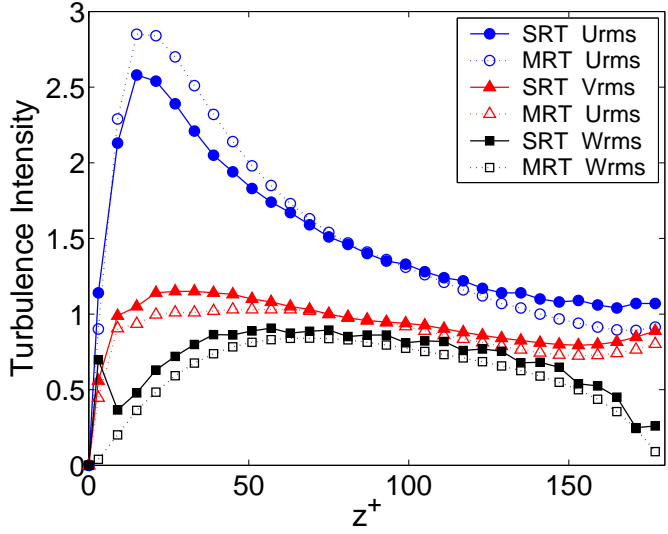}% Here is how to import EPS art
\caption{\label{fig:SRT_MRT_StabilityCompare} Comparison of the
components of root-mean-square (rms) velocity fluctuations
normalized by the wall shear velocity for fully-developed turbulent
channel flow with a free-slip surface at the top for $\mathrm{Re}_{*} = 180$
obtained by GLBE or MRT-LBE (dashed lines with open symbols) and
BGK-LBE or SRT-LBE (solid line with filled symbols) on a coarse
grid.}
\end{figure}

Figure~\ref{fig:SRT_MRT_StabilityCompare} shows the components of
rms turbulent fluctuations obtained by using both the GLBE and the
SRT-LBE at $\nu_0=0.0012$.
%%%%% FIGURE %%%%%
The rms turbulent fluctuations results from the SRT-LBE simulation
show some physically unrealistic behavior, with a large spike in the
wall normal component  near the no-slip wall. Farther out, ripples
which grow as the slip-surface is approached can be seen in both the
wall normal and the streamwise component. That is, spurious
oscillations due to non-hydrodynamic or kinetic modes seem to
strongly interact with fluctuating turbulent motions generated by
the wall, particularly in the wall-normal component, in the case of the
SRT-LBE. In contrast, due to scale separation of relaxation
parameters in the GLBE, the kinetic modes are quickly damped and do not
exhibit such unphysical behavior.

It may be noted that such spurious effects do not seem to manifest
with the SRT-LBE, when fine enough resolution is employed, as was
also noticed, for e.g. in Ref.~\cite{lammers06}. On the other hand, for the
same resolution, if the viscosity is lowered further, the SRT-LBE
becomes unstable. Stable and physically realistic solutions can be
obtained only for viscosity greater than 0.0018 in this particular
case. On the other hand, the GLBE seems to predict correct physical
and smoother behavior for all the components of velocity
fluctuations for viscosity of 0.0012 shown in
Fig.~\ref{fig:SRT_MRT_StabilityCompare} and up to 0.0006 in our
work. For this specific problem we thus obtain enhancement in
stability by a factor of about $3$, which is consistent with the
observations made for other
problems~\cite{dhumieres02,dellar03b,mccracken05,premnath05a,premnath06,yu06}.
Thus, it appears that the GLBE is superior in terms of both physical fidelity
and stability on coarser grid LES simulations of anisotropic and
inhomogeneous turbulent flows, when it is used in lieu of the SRT-LBE.
We will also discuss more on the stability aspects when we discuss
about the other canonical problem considered in this paper.

\subsection{\label{sec:multiblockturbchannelflow}Conservative Multiblock Approach for Local Grid Refinement}
Close to a wall, length scales are very small, requiring a fine
grid to adequately resolve turbulent structures. Use of a grid fine
enough to resolve the wall regions throughout the domain can entail
significant computational cost, and this can be mitigated by
introducing coarser grids farther from the wall, where turbulent
length scales are larger. One approach is to consider using
continuously varying grid resolutions, using a
interpolated-supplemented LBM~\cite{he96} that effectively decouples
particle velocity space represented by the lattice and the
computational grid. However, it is well known that interpolation
could introduce significant numerical dissipation, see for
e.g.~\cite{lallemand00}, which could severely affect the accuracy of
solutions involving turbulent fluctuations, as was confirmed in numerical
experiments during the course of this work. Thus, we consider locally embedded
grid refinement approaches, and in particular their conservative
versions~\cite{chen06,rohde06} that enforce mass and momentum
conservation. Similar zonal embedded approaches have been
successfully employed in computational approaches based on the
solution of filtered NSE for LES of turbulent
flows~\cite{kravchenko96}.

\begin{figure}
\includegraphics[width = 120mm,viewport=0 0 540
270,clip]{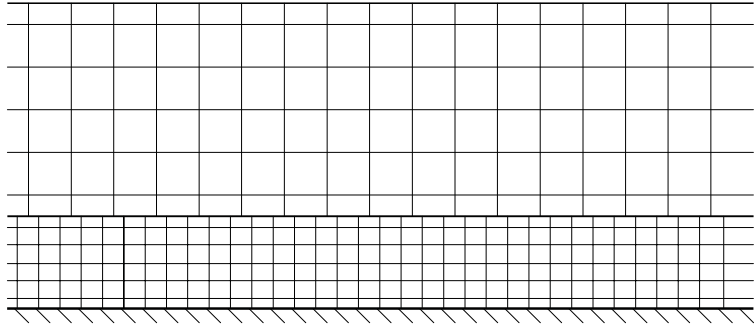}% Here is how to import EPS art
\caption{\label{fig:multiblockschematic} Schematic of conservative
local refinement using multiblock grids with GLBE.}
\end{figure}

Figure~\ref{fig:multiblockschematic} shows a schematic of such a
multiblock approach in which a fine cubic lattice grid is used close
to the wall and a coarser one, again cubic in shape, farther out.
%%%%% FIGURE %%%%%
In order to facilitate the exchange of information at the interface
between the grids, the spacing of the nodes changes by an integer
factor, in this case two.  As well as using different grid sizes,
the two regions use different time steps (time step being
proportional to grid size), and the computational cost required per
unit volume is thus reduced by a factor of 16 in the coarse grid.
Figure~\ref{fig:multiblockschematic} shows a staggered  grid
arrangement, in which nodes on the fine and coarse sides of the
interface are arranged in a manner that facilitates the imposition
of mass and momentum conservation. Different blocks communicate with
each other through the \emph{Coalesce} and the \emph{Explode} steps,
in addition to the standard \emph{stream-and-collide} procedure. The
details are provided in Chen \emph{et~al}.~\cite{chen06} and
Rohde \emph{et~al}.~\cite{rohde06}, and here we very
briefly present the essential elements in what follows. The
\emph{Coalesce} procedure involves summing the particle populations
on the fine nodes to provide new incoming particle populations for
the corresponding coarse nodes. Similarly, the \emph{Explode} step
involves redistributing the populations on the coarse node to the
surrounding fine nodes. These grid-communicating steps used in the
multiblock approach presented in Chen \emph{et~al}.~\cite{chen06} were
incorporated in the GLBE framework in this work.

We performed fully-developed turbulent channel flow at the same
shear Reynolds number as before, i.e. $183.6$, with different
blocks, viz., fine block near the wall and coarse block in the bulk
bounded by top free-slip surface. For the fine grid, we used a
resolution $\Delta^{+}_{fine}=4$ in wall-units (with
$\Delta^{+}_{nw}=2$ due to link-bounce back) and a resolution of
$\Delta^{+}_{coarse}=8$ in wall units for the coarse grid. We used
$256\times 128 \times 17$ grids for the fine block and $128\times 64
\times 17$ for the coarse block, which corresponds to similar aspect
ratios to that used in the earlier simulations. The initial run and
averaging times used were similar to that for the uniform grid case,
viz., $50T^{*}$ and $30T^{*}$, respectively.

\begin{figure}
\includegraphics[width = 100mm,viewport=0 0 540
450,clip]{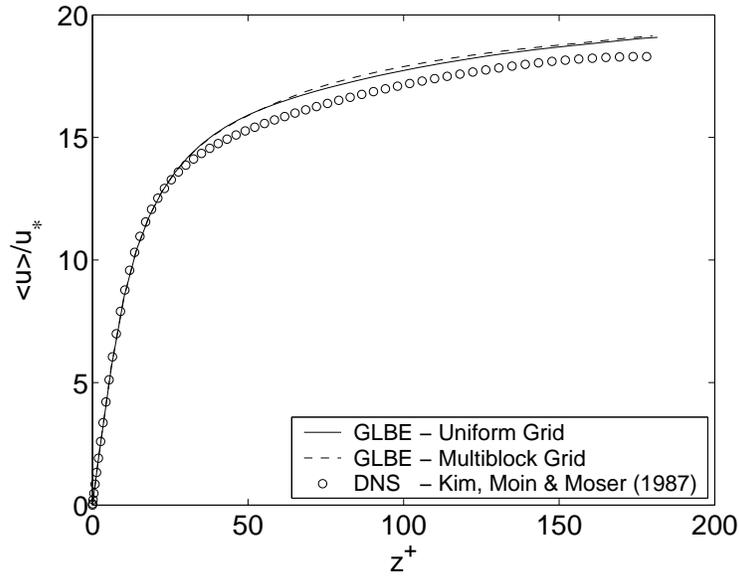}% Here is how to import EPS art
\caption{\label{fig:multiblockmeanvelocity} Mean velocity normalized
by the wall shear velocity for fully-developed turbulent channel
flow at $\mathrm{Re}_{*} = 183.6$. Lines are GLBE results on locally refined
multiblock (broken) and uniform (solid) grids, and symbols are Kim,
Moin and Moser's DNS data (1987)~\cite{kim87}.}
\end{figure}

\begin{figure}
\includegraphics[width = 100mm,viewport=0 0 540
450,clip]{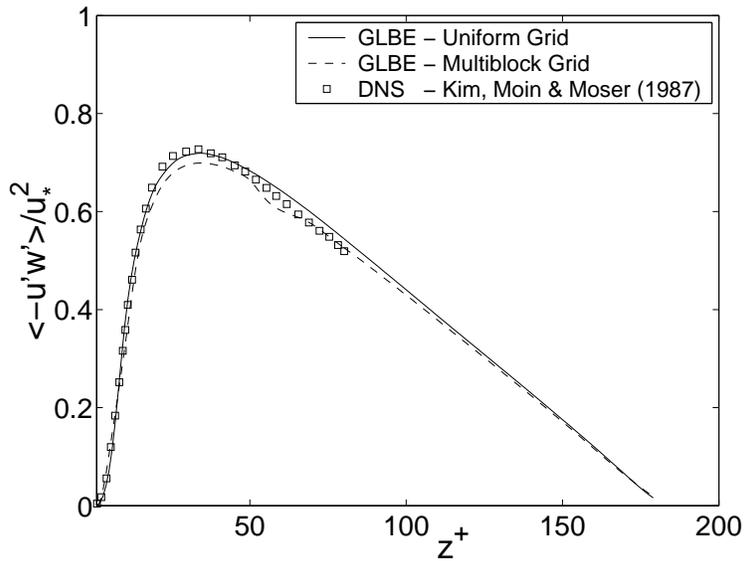}% Here is how to import EPS art
\caption{\label{fig:multiblockreynoldsstress} Reynolds stress
normalized by the wall shear stress for fully-developed turbulent
channel flow at $\mathrm{Re}_{*} = 183.6$. Lines are GLBE results on locally
refined multiblock (broken) and uniform (solid) grids, and symbols
Kim, Moin and Moser's DNS data (1987)~\cite{kim87}.}
\end{figure}

\begin{figure}
\includegraphics[width = 100mm,viewport=0 0 540
450,clip]{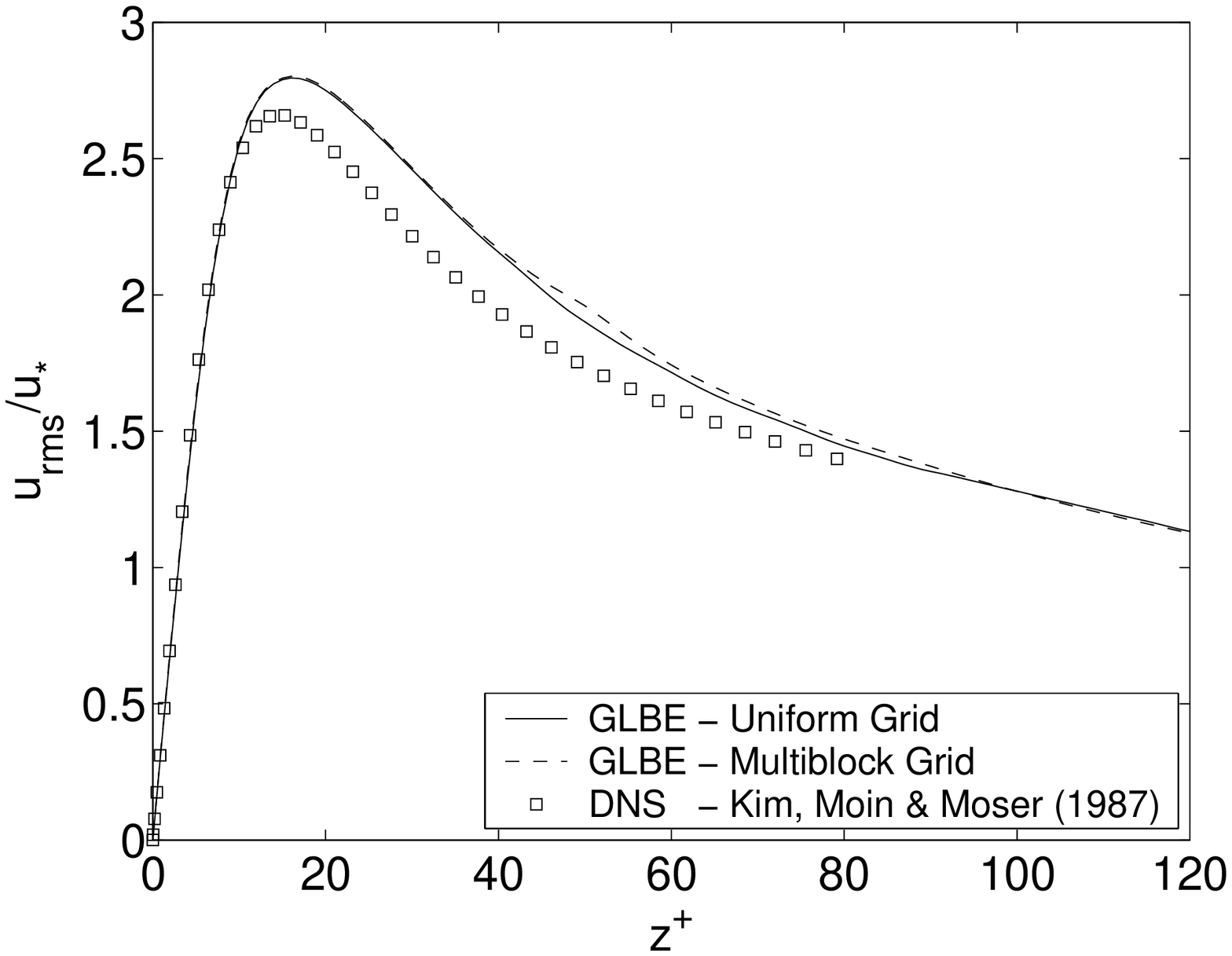}% Here is how to import EPS art
\caption{\label{fig:multiblockurms} Root-mean-square (rms)
streamwise velocity fluctuations normalized by the wall shear
velocity for fully-developed turbulent channel flow at $\mathrm{Re}_{*} =
183.6$. Lines are GLBE results on locally refined multiblock
(broken) and uniform (solid) grids, and symbols are Kim, Moin and
Moser's DNS data (1987)~\cite{kim87}.}
\end{figure}

\begin{figure}
\includegraphics[width = 100mm,viewport=0 0 540
450,clip]{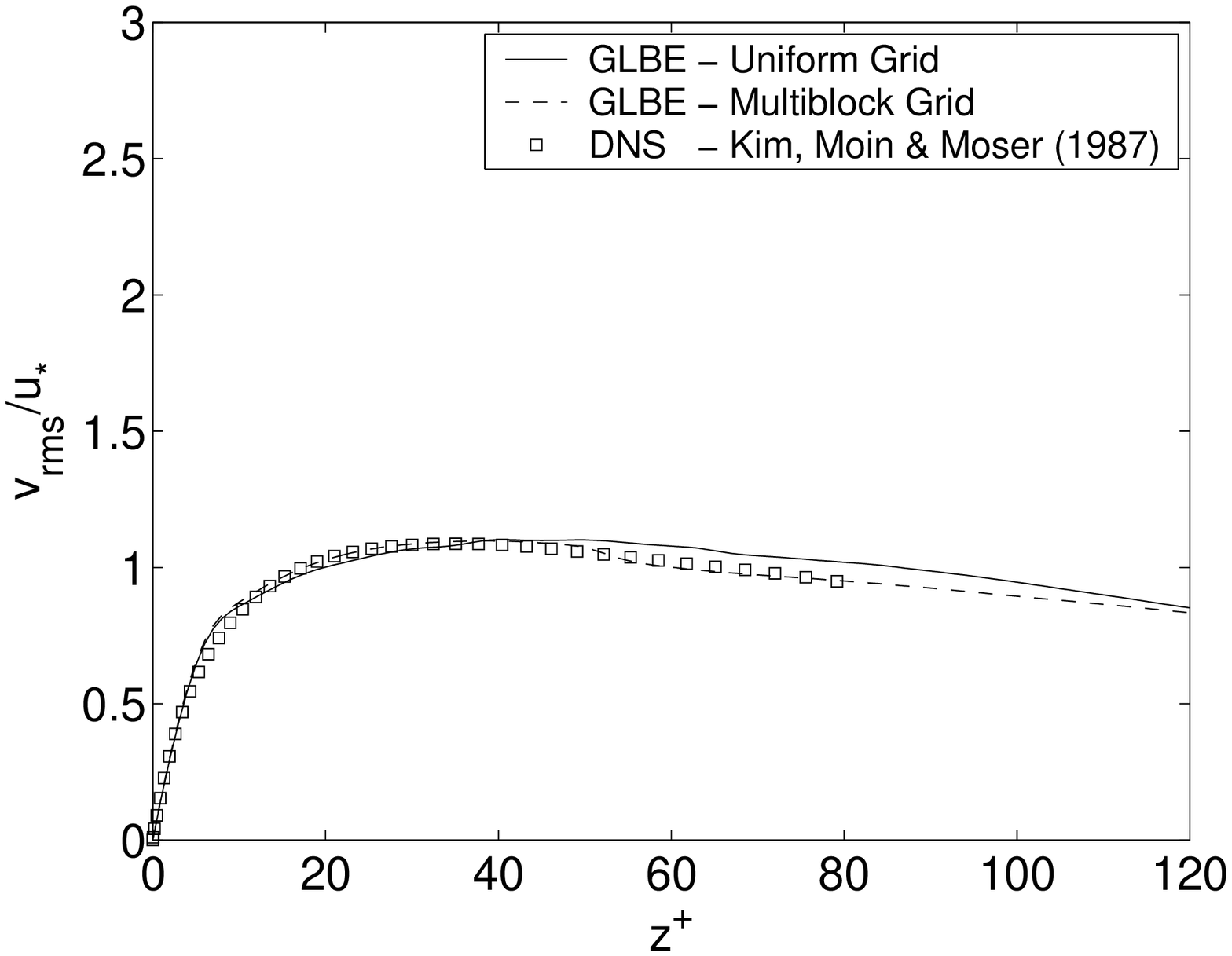}% Here is how to import EPS art
\caption{\label{fig:multiblockvrms} Root-mean-square (rms) spanwise
velocity fluctuations normalized by the wall shear velocity for
fully-developed turbulent channel flow at $\mathrm{Re}_{*} = 183.6$. Lines
are GLBE results on locally refined multiblock (broken) and uniform
(solid) grids, and symbols are Kim, Moin and Moser's DNS data
(1987)~\cite{kim87}.}
\end{figure}

\begin{figure}
\includegraphics[width = 100mm,viewport=0 0 540
450,clip]{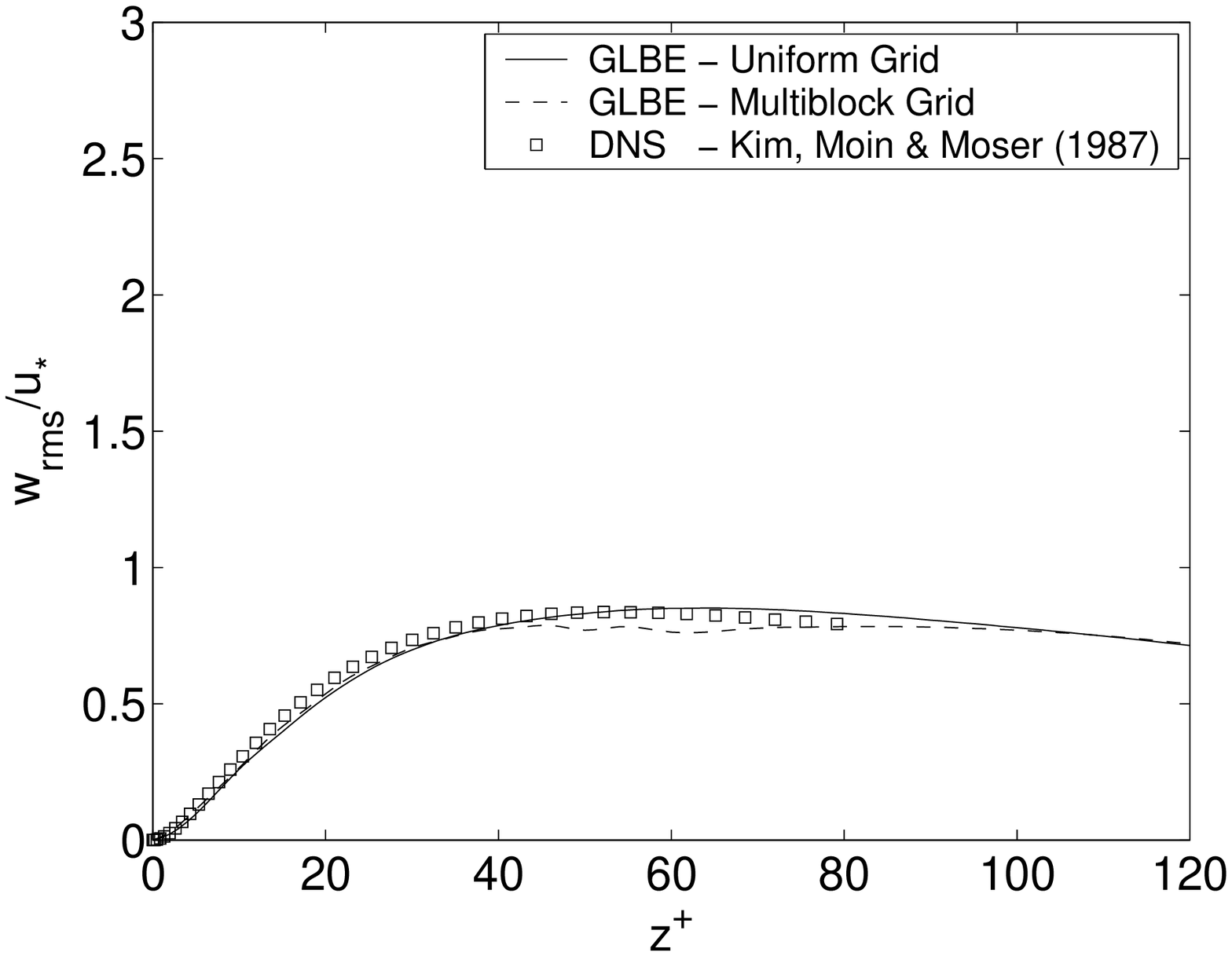}% Here is how to import EPS art
\caption{\label{fig:multiblockwrms} Root-mean-square (rms) wall
normal velocity fluctuations normalized by the wall shear velocity
for fully-developed turbulent channel flow at $\mathrm{Re}_{*} = 183.6$.
Lines are GLBE results on locally refined multiblock (broken) and
uniform (solid) grids, and symbols are Kim, Moin and Moser's DNS
data (1987)~\cite{kim87}.}
\end{figure}

\begin{figure}
\includegraphics[width = 100mm,viewport=0 0 540
450,clip]{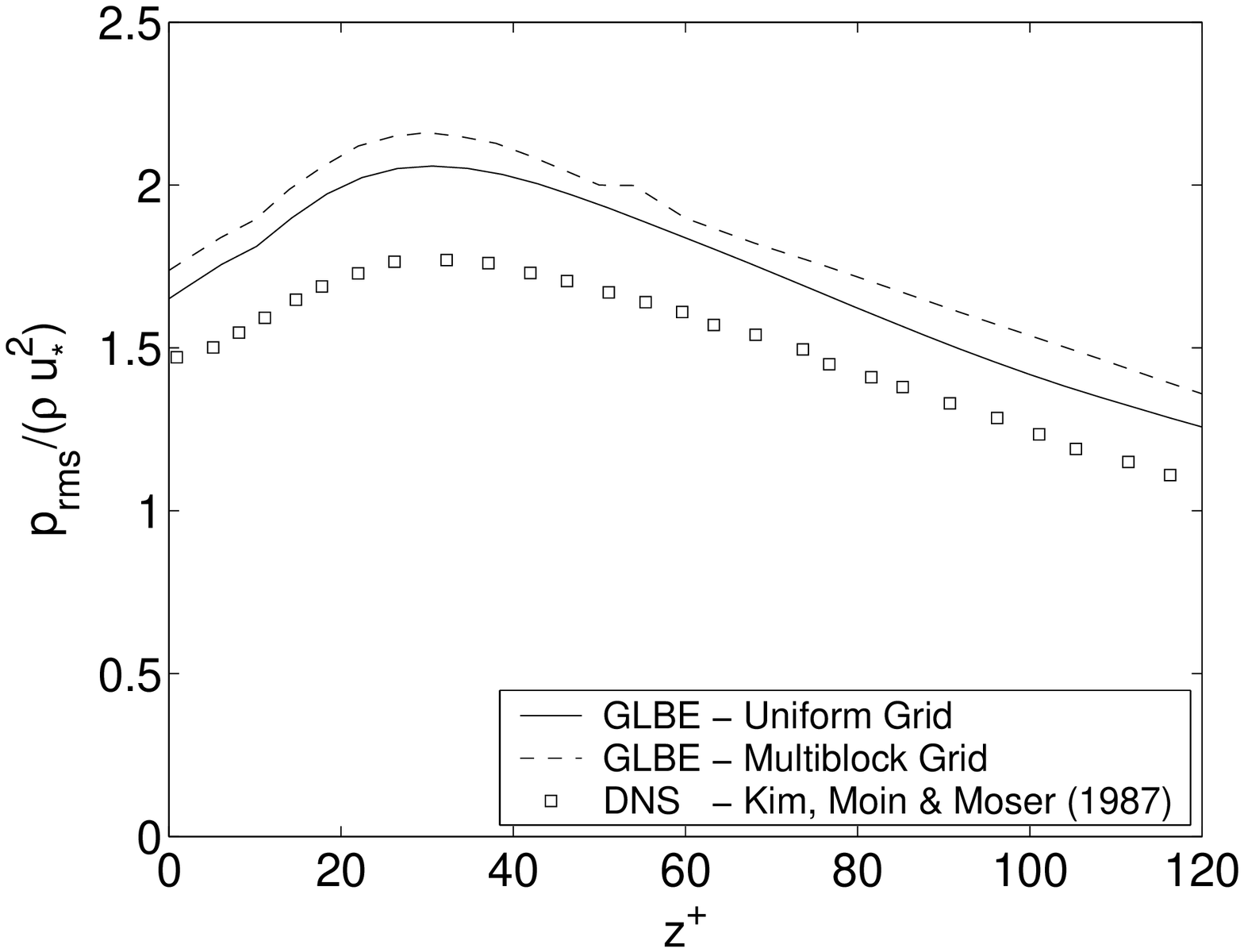}% Here is how to import EPS art
\caption{\label{fig:multiblockprms} Root-mean-square (rms) pressure
fluctuations normalized by the wall shear stress for fully-developed
turbulent channel flow at $\mathrm{Re}_{*} = 183.6$. Lines are GLBE results
on locally refined multiblock (broken) and uniform (solid) grids,
and symbols are Kim, Moin and Moser's DNS data (1987)~\cite{kim87}.}
\end{figure}

The mean velocity and Reynolds stress profiles computed using the GLBE
with locally refined multiblock grids are compared with uniform grid
solution, again computed using GLBE, along with the DNS data~\cite{kim87}, in
Figs.~\ref{fig:multiblockmeanvelocity} and
\ref{fig:multiblockreynoldsstress}, respectively.
%%%%% FIGURE %%%%%
%%%%% FIGURE %%%%%
Generally, good agreement between various simulations can be seen.
Some differences between the DNS data and the LES results based on the
GLBE noticed in these figures are similar to those found in LES
based on filtered NSE. The components of the rms velocity
fluctuations in streamwise, spanwise and wall-normal directions are
presented in Figs.~\ref{fig:multiblockurms},~\ref{fig:multiblockvrms}
and~\ref{fig:multiblockwrms}, respectively.
%%%%% FIGURE %%%%%
%%%%% FIGURE %%%%%
%%%%% FIGURE %%%%%
Again, the multiblock GLBE based LES results are fairly in good
agreement with the uniform grid GLBE as well as the DNS data for various
components of velocity fluctuations. It is found that the velocity fluctuations
and Reynolds stress are somewhat sensitive to
numerical artifacts arising near grid-transition regions, i.e. at
the interface between fine and coarse grid blocks, where they are
slightly damped. Similar features have been noted in
Ref.~\cite{rohde06} when the multiblock approach is employed for
computation of certain classes of flows, having flow components
normal to grid interfaces. On the other hand, when the rms pressure
fluctuations computed using the multiblock GLBE are compared with
uniform grid results, there is a slight overprediction by the
former, plausibly due to added compressibility effects with the use
of multiblock grids (see Fig.~\ref{fig:multiblockprms}).
%%%%% FIGURE %%%%%
%It appears that multiblock GLBE with forcing term is a reliable and
%accurate tool, with enhanced stability, for wall-bounded turbulent
%flows.

\subsection{\label{sec:parallelscalability}Parallel Scalability}
\begin{figure}
\includegraphics[width = 120mm,viewport=0 0 720
570,clip]{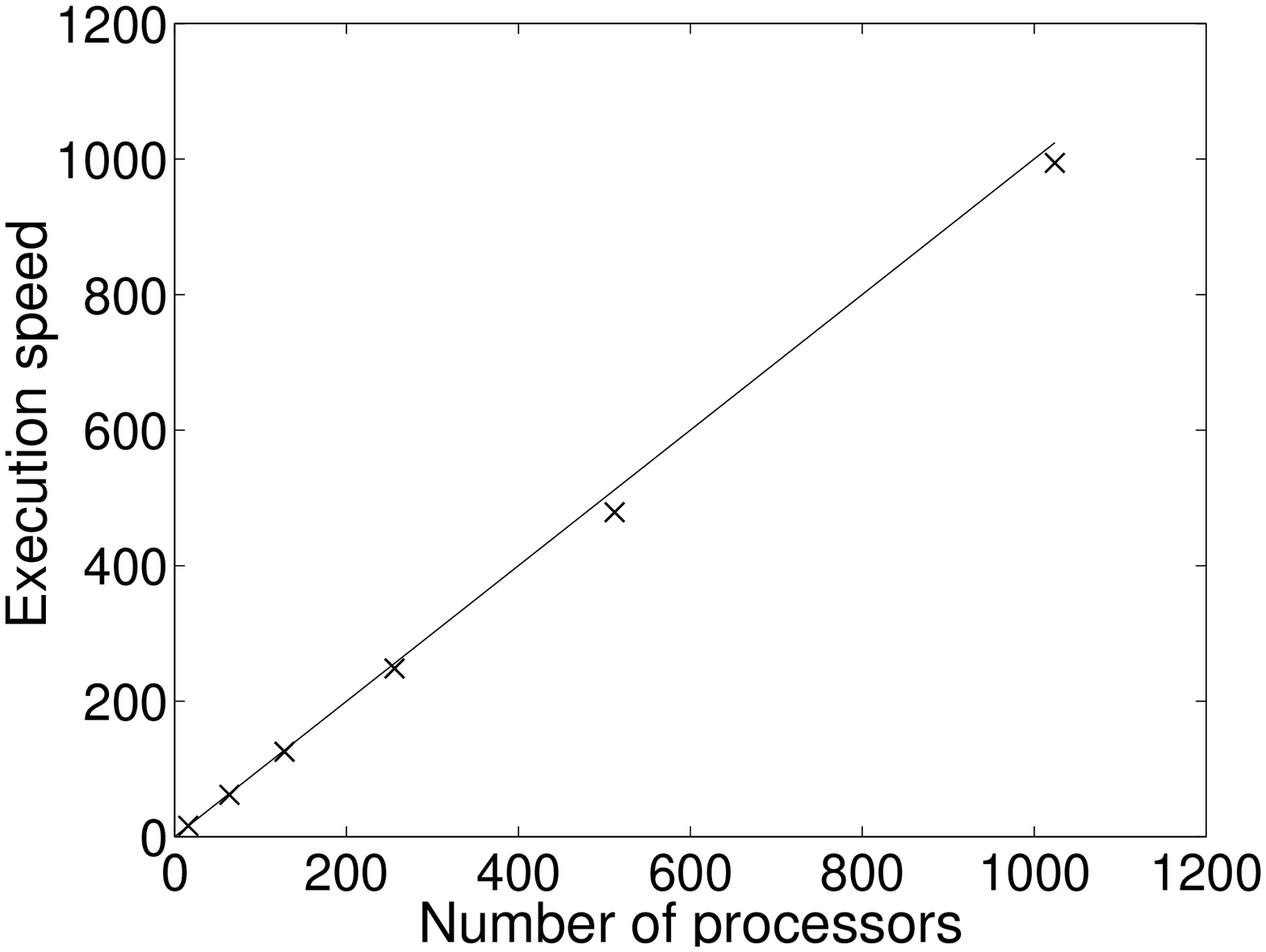}% Here is how to import EPS art
\caption{\label{fig:parallelspeedup1} Parallel scalability of GLBE
for turbulence simulation on a parallel cluster, Seaborg, at
Department of Energy's NERSC. Symbols are GLBE performance and line
is the ideal linear speed-up.}
\end{figure}

One of the main advantages of the LBM is its natural amenability for
implementation on parallel computers. The code implementation of
GLBE with forcing term was parallelized using the Message Passing
Interface library through a domain decomposition strategy that
exploits the local and explicit nature of the approach. It was
tested for parallel scalability on a large parallel cluster known as
\emph{Seaborg} located at U.S. Department of Energy's NERSC center.
In these tests, the size of each subdomain was held constant at
$20\times 256\times 256$ ($1.3$ million grid nodes), per processor
for wall-bounded turbulence simulations. The speed-up factors
obtained for up to 1024 processors are shown in
Fig.~\ref{fig:parallelspeedup1}.
%%%%% FIGURE %%%%%
It is evident that near-linear scaling can be obtained on massively
parallel clusters and thus it appears that the GLBE approach is
well-suited for large-scale turbulent flow simulations.

\section{\label{sec:3dcavityflow}Three-Dimensional Flow in a Cubical Cavity}
Let us now consider another wall-bounded flow problem, viz., 3D flow
in a cubical cavity driven by its top lid, and its computation using
the GLBE. Although the geometry is simple, it is characterized by
richness in fluid flow physics as there are no homogeneous
directions and the presence of walls on all sides profoundly
modifies the flow behavior. Features such as multiple
counter-rotating recirculating regions at the corners,
Taylor--G\"{o}rtler-type vortices, bifurcations in flows and
transition to turbulence may manifest themselves depending on the
Reynolds number~\cite{shankar00}. Generally, when the Reynolds
number $\mathrm{Re}$ based on the cavity side length is less than 2000, the
flow field is laminar, and flow instabilities manifest themselves
near the downstream corner eddy when $\mathrm{Re}$ is between 2000 and 3000.
As $\mathrm{Re}$ increases, turbulence is generated near the cavity walls, with the
flow near the downstream corner eddy becoming fully turbulent when
$\mathrm{Re}{\lower5pt\hbox{$>$}\atop\raise1pt\hbox{$\sim$}} 10,000$. Due to various states exhibited by the flow at
higher $\mathrm{Re}$ it is a very challenging problem to study, in particular
in obtaining computational results as it requires accurate methods
with long averaging times. Measurements from experiments in cubical
cavity are available for $\mathrm{Re}=10,000$ in Prasad and Koseff~\cite{prasad89}, while
pseudo-spectral DNS and spectral element LES were performed more
recently at $\mathrm{Re}=12,000$ by Leriche and Gavrilakis~\cite{leriche00}
and Bouffanais \emph{et~al}.~\cite{bouffanais07}, respectively.
In the context of LBM, d'Humi\`{e}res \emph{et~al}.~\cite{dhumieres02}
performed simulations of 3D flow in a diagonally driven cavity, in the
laminar and transition regime, i.e. $\mathrm{Re}\leq 4000$. The focus here is
to perform GLBE simulations at a higher $\mathrm{Re}$ of $12,000$ and
compare results with available prior computational
results~\cite{leriche00,bouffanais07} and experimental
data~\cite{prasad89}. In addition, we also compare numerical stability
of the GLBE and the SRT-LBE for this problem at higher $\mathrm{Re}$ range.

\begin{figure}
\includegraphics[width = 130mm,viewport=90 100 550
470,clip]{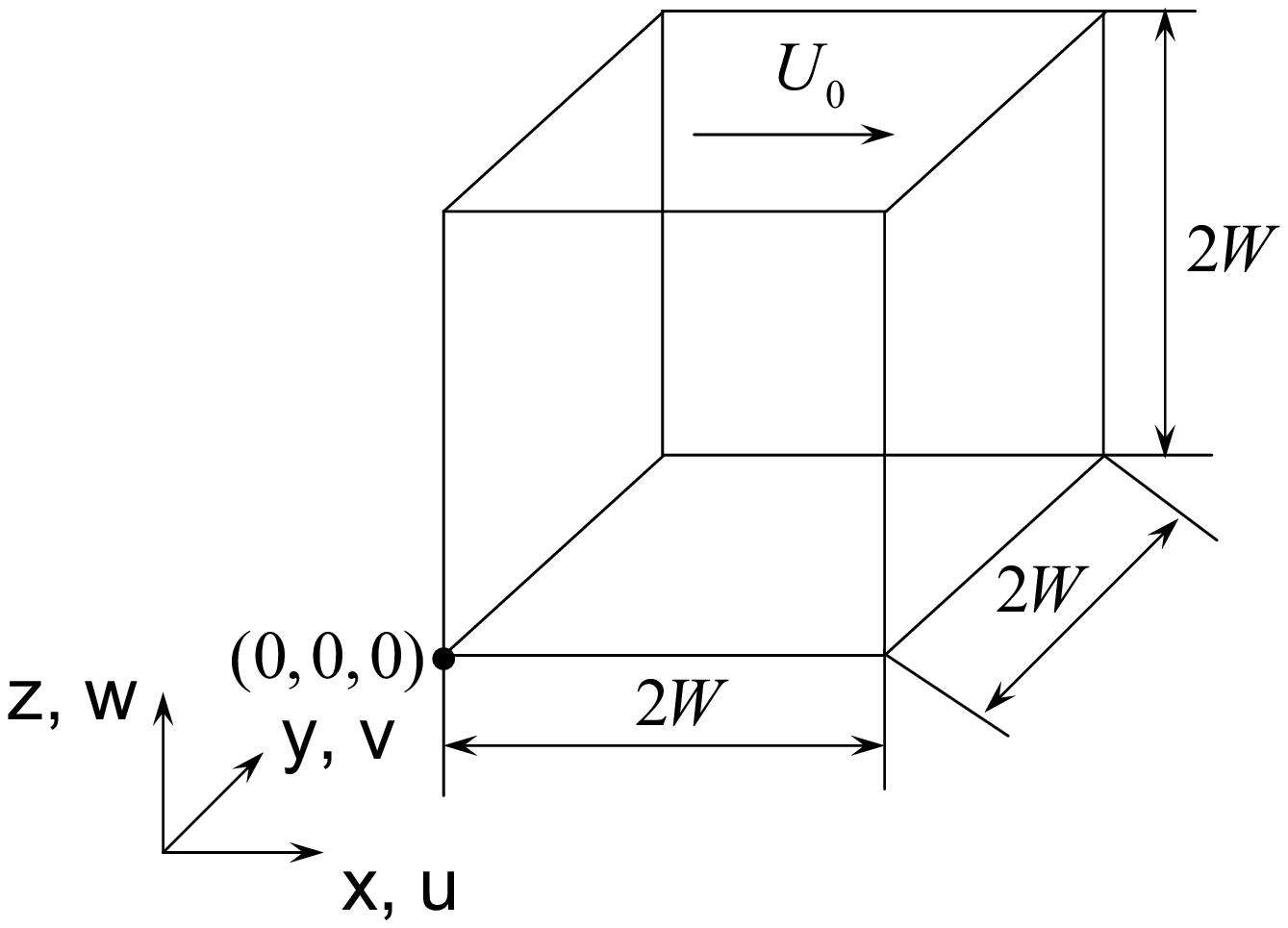}% Here is how to import EPS art
\caption{\label{fig:drivencavity} Schematic arrangement of
computational domain for   LES of flow in a three-dimensional
cubical cavity of half-width $W$ driven by its top lid with velocity
$U_{0}$.}
\end{figure}

The computational conditions that we considered are as follows. The
schematic of the 3D flow in a cubic cavity of side length $2W$ is
shown in Fig.~\ref{fig:drivencavity} with a coordinate system, in
which flow is driven by the top lid with velocity $U_{0}$.
%%%%% FIGURE %%%%%
%%%%% FIGURE %%%%%
The Reynolds number used in our computations $\mathrm{Re}=U_0
2W/\nu_0=12,000$ is achieved by setting the lid velocity to be
$U_0=0.12$, with a viscosity of $\nu_0=0.00128$ on a relatively fine
uniform grid with $128^3$ lattice nodes. A note regarding the choice of
the lid velocity is in order. For a given $\mathrm{Re}$, when the fluid viscosity
is chosen based on relaxation parameter so as to maintain numerical
stability, the choice of $U_0$ influences the number of grid nodes needed
to resolve \emph{each} side of the 3D cubic cavity. That is, any reduction of
$U_0$ by a factor $k$ will increase the total number of grid nodes by $k^3$.
On the other hand, since the GLBE is a weakly compressible computational
approach, the Mach number $\mathrm{Ma}$ ($=U_0/c_s$ where $c_s=1/\sqrt{3}$)
should be small. Thus, the value of $U_0$ is chosen as a compromise
between satisfying the weakly compressible condition and resolution requirements
so as the obtain an acceptable level of solution accuracy.
In this work, as found later, the choice of $U_0=0.12$ and a resolution with
$128^3$ grid nodes yields reasonably good accuracy.

Now, imposing a constant lid velocity profile on the top lid leads to edge and
corner singularities and can significantly affect the stability, convergence and
accuracy of simulations at such high
$\mathrm{Re}$~\cite{leriche00,bouffanais07}. In reality, there is a velocity
distribution at the top lid, whose precise form is not known.
Following Leriche and Gavrilakis~\cite{leriche00} as well as
Bouffanais \emph{et~al}.~\cite{bouffanais07}, we set the
following velocity profile for the lid:
\begin{equation}
u_{lid}(x,y)=u_0\left[1-\left(\frac{x-W}{W}\right)^{18}
\right]^2\left[1-\left(\frac{y-W}{W}\right)^{18} \right]^2.
\end{equation}
It was found that the flow field in the cavity is not overly
sensitive to the lid velocity profile, when such higher-order
polynomial distributions are used~\cite{leriche00,bouffanais07}. The
mean value of this velocity profile is $U_m \approx 0.85U_0$, with
over $75\%$ area of the lid has a velocity above $U_m$ and the
corresponding Reynolds number on the mean velocity is $10,200$. In the
GLBE, the velocity boundary condition at the lid is provided by
setting the distribution function of incoming populations
corresponding to $\overrightarrow{e}_{\overline{\alpha}}$ through an
momentum-augmented bounce back as follows~\cite{bouzidi01}:
%\begin{equation}
%f_{\alpha}=w_{\alpha}\rho_0\frac{\overrightarrow{e_{\alpha}}\cdot\overrightarrow{u}_{0}}{c_s^2}
%\end{equation}
\begin{equation}
f_{\overline{\alpha}}=f_{\alpha}+2w_{\alpha}\rho_0\frac{\overrightarrow{e_{\overline{\alpha}}}\cdot\overrightarrow{u}_{lid}}{c_s^2}
\end{equation}
where $\overrightarrow{e}_{\overline{\alpha}}=-\overrightarrow{e}_{\alpha}$. No-slip zero velocity
boundary conditions based on bounce back approach are set for all the
other walls. A statistically stationary state of the flow field is
obtained after running for $500T^{*}$ and then collecting statistics
at each grid node averaged over a period of $150T^{*}$, where the
characteristic time is $T^{*}=2W/u_0$.

\begin{figure}
\includegraphics[width = 100mm]{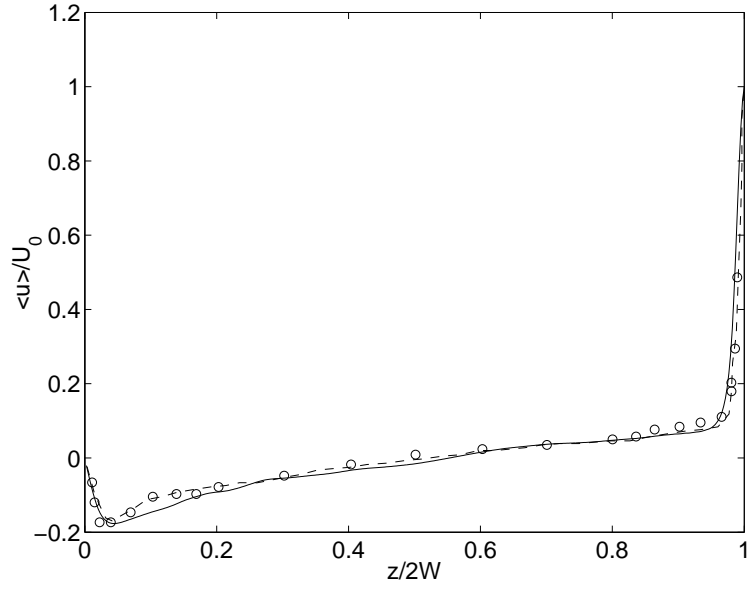}% Here is how to import EPS art
\caption{\label{fig:umeanz} Mean velocity $<u>$ on the centerline
$x=W, y=W$ obtained in LES using GLBE (solid line) compared with DNS
of Leriche and Gavrilakis (2000)~\cite{leriche00} (dashed line) and
experimental data of Prasad and Koseff (1989)~\cite{prasad89}
(circles).}
\end{figure}

\begin{figure}
\includegraphics[width = 100mm]{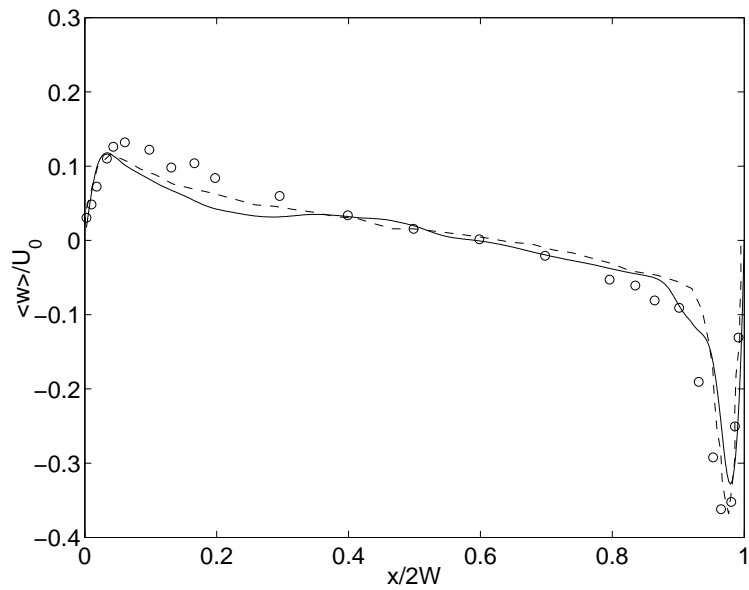}% Here is how to import EPS art
\caption{\label{fig:umeanx} Mean velocity $<w>$ on the centerline
$y=W, z=W$ obtained in LES using GLBE (solid line) compared with DNS
of Leriche and Gavrilakis (2000)~\cite{leriche00} (dashed line) and
experimental data of Prasad and Koseff (1989)~\cite{prasad89}
(circles).}
\end{figure}

Figures~\ref{fig:umeanz} and~\ref{fig:umeanx} show the computed
first-order statistics, viz., the mean velocity profile on two of
the cavity centerlines along with the other available data for
comparison.
%%%%% FIGURE %%%%%
%%%%% FIGURE %%%%%
%%%%% FIGURE %%%%%
%%%%% FIGURE %%%%%
It is seen that GLBE solution is in reasonable agreement with the
DNS~\cite{leriche00} and experimental data~\cite{prasad89}.

\begin{figure}
\includegraphics[width = 100mm]{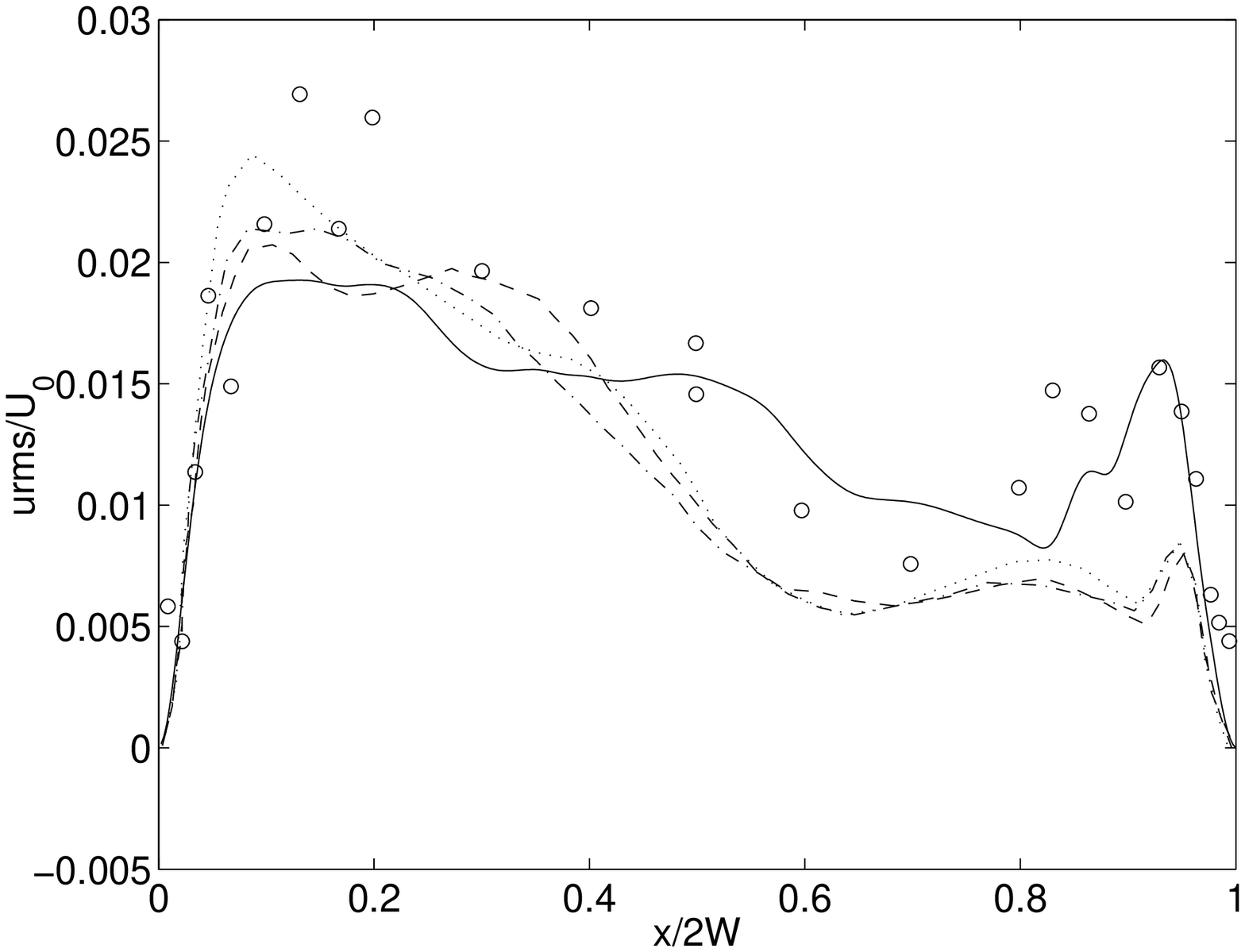}% Here is how to import EPS art
\caption{\label{fig:urmsx} Root mean square (rms) velocity
fluctuations $urms$ on the centerline $y=W, z=W$ obtained in LES
using GLBE (solid line) compared with DNS of Leriche and Gavrilakis
(2000)~\cite{leriche00} (dashed line), LES with NSE of Bouffanais et~al.
(2007)~\cite{bouffanais07} based on dynamic model (dotted line)
and dynamic mixed model (dot-dashed line) and experimental data of
Prasad and Koseff (1989)~\cite{prasad89} (circles).}
\end{figure}

\begin{figure}
\includegraphics[width = 100mm]{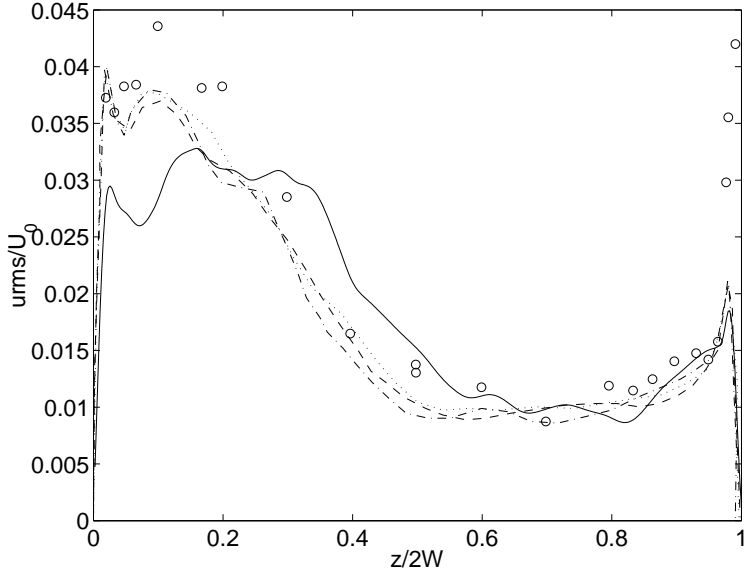}% Here is how to import EPS art
\caption{\label{fig:urmsz} Root mean square (rms) velocity
fluctuations $urms$ on the centerline $x=W, y=W$ obtained in LES
using GLBE (solid line) compared with DNS of Leriche and Gavrilakis
(2000)~\cite{leriche00} (dashed line), LES with NSE of Bouffanais et~al.
(2007)~\cite{bouffanais07} based on dynamic model (dotted line)
and dynamic mixed model (dot-dashed line) and experimental data of
Prasad and Koseff (1989)~\cite{prasad89} (circles).}
\end{figure}

\begin{figure}
\includegraphics[width = 100mm]{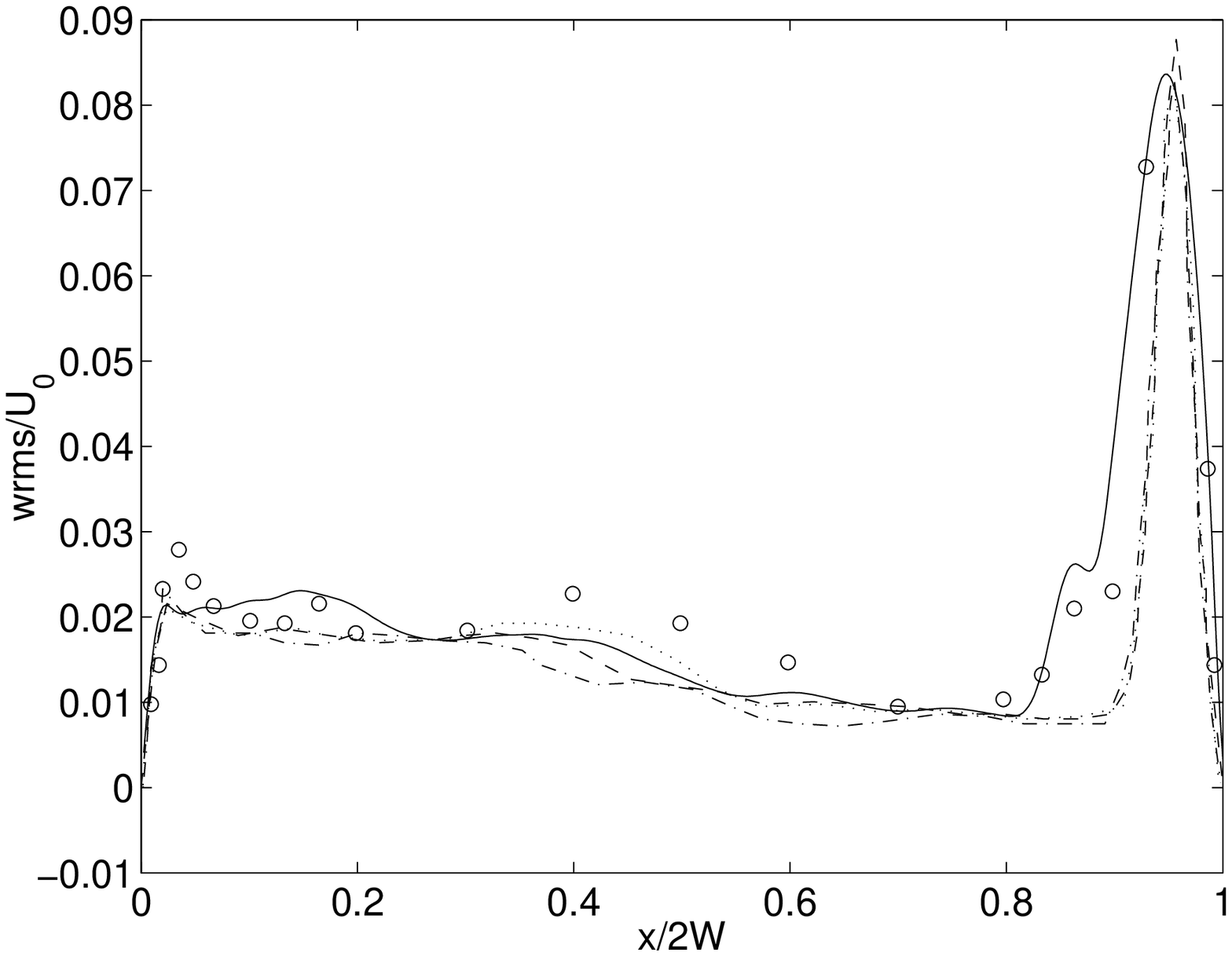}% Here is how to import EPS art
\caption{\label{fig:wrmsx} Root mean square (rms) velocity
fluctuations $wrms$ on the centerline $y=W, z=W$ obtained in LES
using GLBE (solid line) compared with DNS of Leriche and Gavrilakis
(2000)~\cite{leriche00}(dashed line), LES with NSE of Bouffanais et~al.
(2007)~\cite{bouffanais07} based on dynamic model (dotted line)
and dynamic mixed model (dot-dashed line) and experimental data of
Prasad and Koseff (1989)~\cite{prasad89} (circles).}
\end{figure}

\begin{figure}
\includegraphics[width = 100mm]{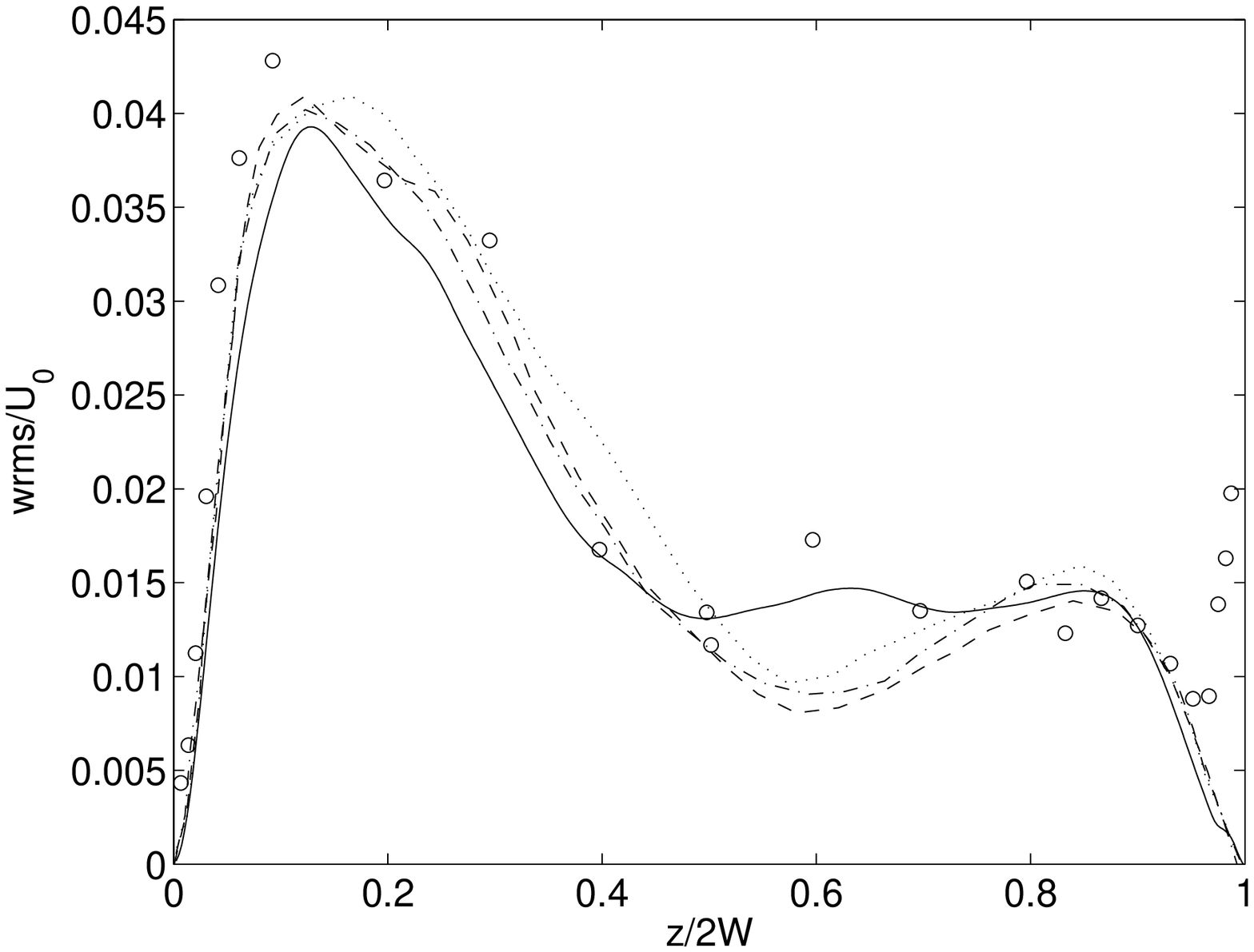}% Here is how to import EPS art
\caption{\label{fig:wrmsz} Root mean square (rms) velocity
fluctuations $wrms$ on the centerline $x=W, y=W$ obtained in LES
using GLBE (solid line) compared with DNS of Leriche and Gavrilakis
(2000)~\cite{leriche00}(dashed line), LES with NSE of Bouffanais et~al.
 (2007)~\cite{bouffanais07} based on dynamic model (dotted line)
and dynamic mixed model (dot-dashed line) and experimental data of
Prasad and Koseff (1989)~\cite{prasad89} (circles).}
\end{figure}

\begin{figure}
\includegraphics[width = 100mm]{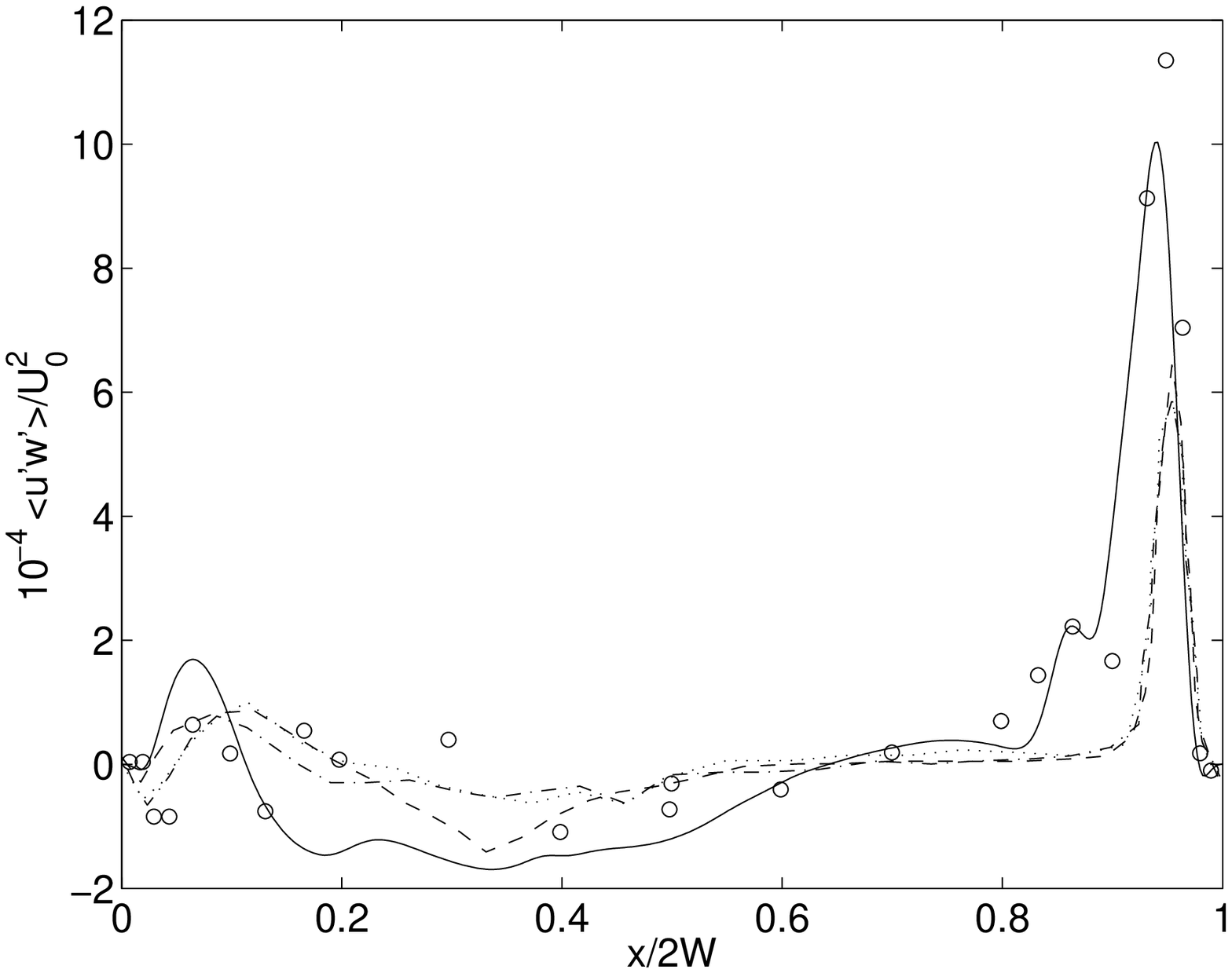}% Here is how to import EPS art
\caption{\label{fig:reystressx} Reynolds stress  $<u'w'>$ on the
centerline $y=W, z=W$ obtained in LES using GLBE (solid line)
compared with DNS of Leriche and Gavrilakis (2000)~\cite{leriche00}
(dashed line), LES with NSE of Bouffanais et~al.
(2007)~\cite{bouffanais07} based on dynamic model (dotted line) and
dynamic mixed model (dot-dashed line) and experimental data of
Prasad and Koseff (1989)~\cite{prasad89} (circles).}
\end{figure}

\begin{figure}
\includegraphics[width = 100mm]{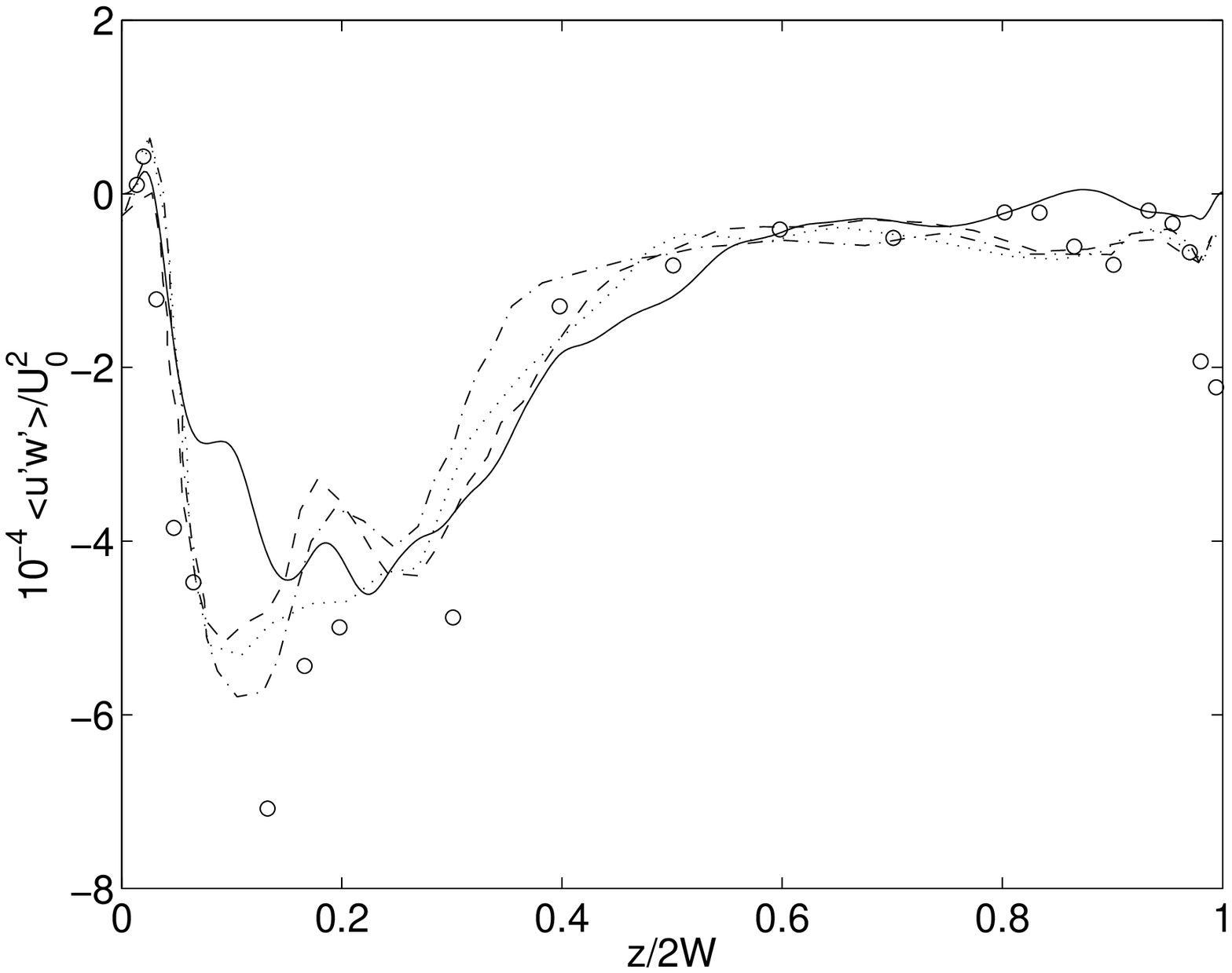}% Here is how to import EPS art
\caption{\label{fig:reystressz} Reynolds stress  $<u'w'>$ on the
centerline $x=W, y=W$ obtained in LES using GLBE (solid line)
compared with DNS of Leriche and Gavrilakis (2000)~\cite{leriche00}
(dashed line), LES with NSE of Bouffanais et~al.
(2007)~\cite{bouffanais07} based on dynamic model (dotted line) and
dynamic mixed model (dot-dashed line) and experimental data of
Prasad and Koseff (1989)~\cite{prasad89} (circles).}
\end{figure}

In general, as discussed in Ref.~\cite{leriche00}, momentum transfer
from the lid creates a region of high pressure in the upper corner
of the downstream wall as the flow has to change direction,
dissipating part of its energy. The flow then convects downwards
along the downstream wall like an unsteady wall jet, which separates
from the wall near the mid-section of the wall and leading to two
elliptical jets. They subsequently impinge on the bottom cavity wall
and generate turbulence, which is convected away by the central and main
vortex. The second-order statistics of the fluctuating flow field
provide an indication of the turbulent activity.
Figures~\ref{fig:urmsx} and~\ref{fig:urmsz} provide the rms velocity
fluctuations along the direction parallel to the lid motion, i.e.
$u_{rms}$ on the centerlines $y=W, z=W$ and $x=W, y=W$,
respectively,
%%%%% FIGURE %%%%%
%%%%% FIGURE %%%%%
%%%%% FIGURE %%%%%
%%%%% FIGURE %%%%%
%\clearpage
while Figs.~\ref{fig:wrmsx} and~\ref{fig:wrmsz} provide the rms
velocity fluctuations along direction normal to the lid motion i.e.
$w_{rms}$ on the centerlines $y=W, z=W$ and $x=W, y=W$,
respectively.
%%%%% FIGURE %%%%%
%%%%% FIGURE %%%%%
%%%%% FIGURE %%%%%
%%%%% FIGURE %%%%%
In addition, the components of the Reynolds stress $<u'w'>$ on the
centerlines $y=W, z=W$ and $x=W, y=W$ are provided in
Figs.~\ref{fig:reystressx} and~\ref{fig:reystressz}, respectively.
%%%%% FIGURE %%%%%
%%%% FIGURE %%%%%
%%%%% FIGURE %%%%%
%%%%% FIGURE %%%%%
Firstly, these results indeed show that turbulence is generated
along cavity walls. In particular, the turbulent fluctuations are
about an order of magnitude larger near the downstream wall than
near the upstream wall. Moreover, the fluctuations are the largest
along the bottom wall. These seem to be consistent with the
description of the features of the fluid motion in the cavity, as
elucidated in the DNS~\cite{leriche00}. Although there is some deviation
in the peaks of the fluctuations when compared with other data,
considering that DNS and LES considered here are based on approaches
using higher-order spectral methods~\cite{leriche00,bouffanais07},
the GLBE computations, in general, compare reasonably well with
them, which are very encouraging.

Some differences observed between computed solutions, including
those from DNS and LES~\cite{leriche00,bouffanais07}, and the
experimental data could be attributed to differences in Reynolds
number as well as the averaging times used. For example, the magnitude of
the peak value of the near-wall Reynolds stress in Fig.~\ref{fig:reystressx}
is influenced by the length of the time interval over which averaging is
performed. In this work, we have chosen the time period of averaging ($150T^{*}$)
from the sampling period used in experiments~\cite{prasad89}. In effect, our results
are closer to these data. However, it should be noted that prior computations~\cite{leriche00,bouffanais07}
found that the peak value of Reynolds stress is conditioned by rare events, which occur on
time intervals of approximately $80T^{*}$. Hence, the averaging period
in Refs.~\cite{leriche00,bouffanais07} is chosen such that the rare events, which tend to suppress fluctuations,
are sampled many times, which accounts for the difference between
experiments~\cite{prasad89} (including this work) and prior computations~\cite{leriche00,bouffanais07}.

Moreover, a note regarding the influence of the choice of the SGS turbulence model on the turbulence statistics
results is in order. Bouffanais \emph{et~al}.~\cite{bouffanais07}, who employed dynamic SGS models in their
computations yielding high fidelity results in excellent agreement with DNS, also reported preliminary results
without employing a SGS model, i.e. unresolved DNS, and with a constant Smagorinsky SGS model. They found that the
unresolved DNS is not even qualitatively correct for this problem and the use of constant Smagorinsky SGS model
resulted in improved predictions, but still not fully consistent with the resolved DNS and their results with
using dynamic models. However, unlike Bouffanais \emph{et~al}.~\cite{bouffanais07}, in this work,
we have used van Driest wall damping function~\cite{vandriest56} in conjunction with the constant
Smagorinsky SGS model, which appears to make considerable difference in further improving the
quality of the results. It appears that the use of wall damping function, which accounts for reduction of
near-wall turbulent length scales, yields results that are significantly closer and more consistent than
without the use of such a damping function. This is also consistent with our recent observation~\cite{premnath08a}
that the use of a wall damping function with constant Smagorinsky SGS model results in better agreement with
the dynamic SGS model results for LES of turbulent channel flow than without it. It should, however, also be noted that
while the results obtained with the use of damping function are generally better, they are in some cases
somewhat underpredicted near walls, e.g. Fig.~\ref{fig:urmsz}, and the peaks
in some other cases are broader, e.g. Fig.~\ref{fig:wrmsx}. While the motivation for the use
of damping function is simplicity and computational efficiency, it is, of course, desirable to employ dynamic
SGS models in the LBM framework~\cite{premnath08a} for further improvements.

\subsection{\label{sec:stability3dcavityflow}Numerical Stability}
\begin{figure}
\includegraphics[width = 100mm,viewport=0 100 550
480,clip,angle=0]{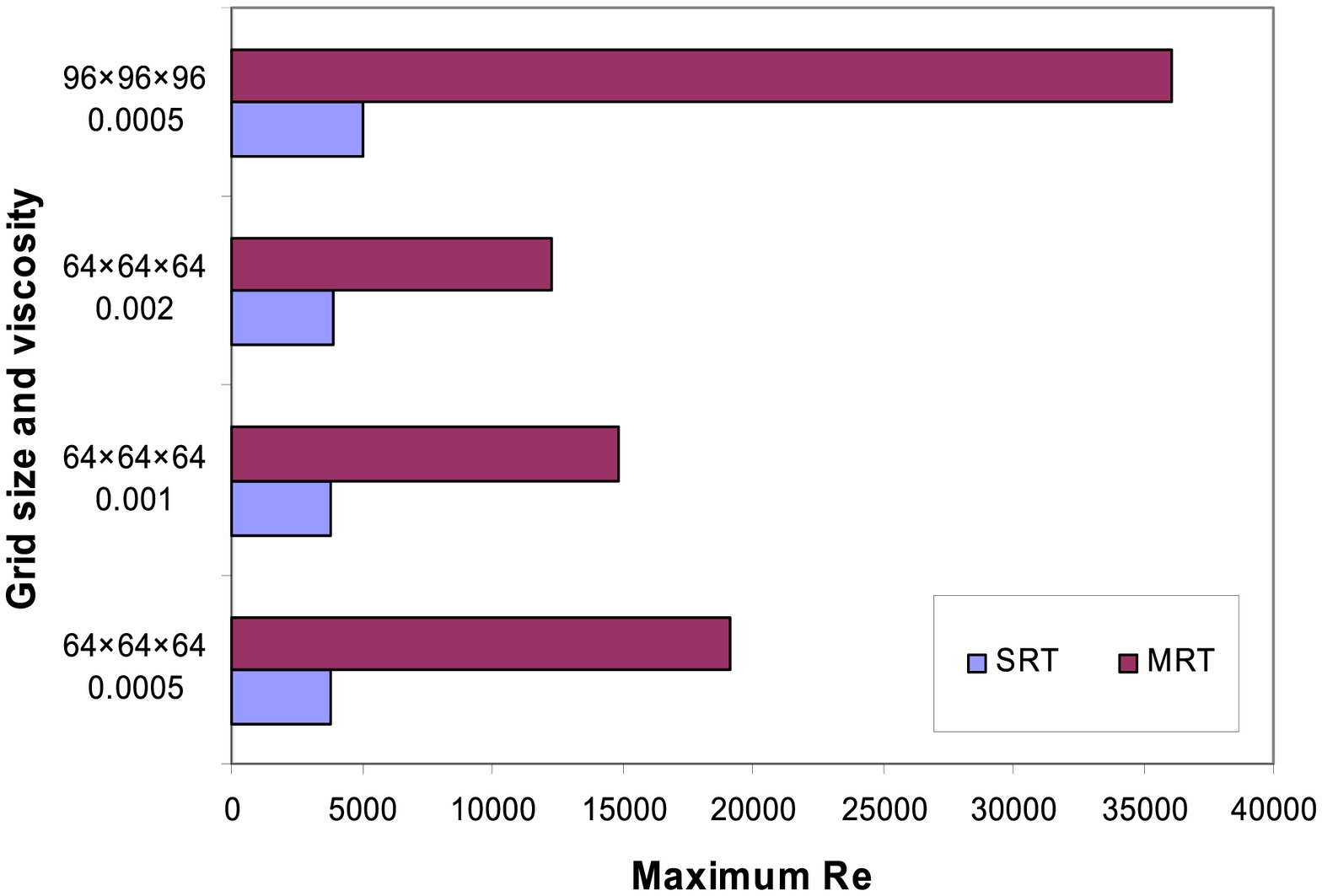}% Here is how to import EPS art
\caption{\label{fig:stabilitycavity} (Color online) Comparison of numerical
stability characteristics of GLBE with SRT-LBE for 3D cavity flows:
Maximum attainable Reynolds number for a given resolution and shear
viscosity.}
\end{figure}

We will now make direct comparisons of stability characteristics of
the GLBE with the SRT-LBE for 3D cavity flow simulations at higher Reynolds
number ranges, complementing an earlier study~\cite{dhumieres02}. In
both approaches, for a given grid resolution, the shear viscosity
was fixed and the lid velocity was increased gradually until the
computation became unstable. Figure~\ref{fig:stabilitycavity} shows
the maximum Reynolds number that could be attained before the
computations became unstable. Results are provided for different
grid resolutions and viscosities for both the approaches.
%\begin{figure}
%\includegraphics[width = 100mm,viewport=0 0 460
%640,clip,angle=270]{Stability_Cavity_Chart_30}% Here is how to import EPS art
%\caption{\label{fig:stabilitycavity} (Color online) Comparison of numerical
%stability characteristics of GLBE with SRT-LBE for 3D cavity flows:
%Maximum attainable Reynolds number for a given resolution and shear
%viscosity.}
%\end{figure}
%%%%% FIGURE %%%%%
The superior stability characteristics of the GLBE are evident for this
wall-bounded turbulent flow problem. The GLBE computations can reach
Reynolds numbers that are several times higher than that of the SRT-LBE,
typically by a factor of $3$ or $4$ and sometimes even about as high
as an order magnitude. Indeed, the SRT-LBE became unstable and unable to
simulate the $\mathrm{Re}=12,000$ case considered above with a resolution of
$128^3$ using the GLBE.

\clearpage

\section{\label{sec:summary}Summary and Conclusions}
A generalized lattice Boltzmann equation (GLBE)
with forcing term, which uses multiple relaxation times, for
eddy-capturing computations of wall-bounded turbulent flows that are
characterized by statistical anisotropy and inhomogeneity is discussed.
Standard Smagorinsky eddy viscosity model is used to represent
SGS turbulence effects, which is modified by the van Driest damping function
to account for reduction of turbulent length scales near walls. Second-order and
effectively time-explicit source terms, which represent general forms of non-uniform
external forces that drive or modulate the character of
turbulent flow, are projected onto the natural moment space of GLBE
in this formulation. In this framework, the strain tensor used in the
SGS model is related to the non-equilibrium moments and the forcing terms in moment space.
Furthermore, local grid refinement using a conservative multiblock approach
is used to coarsen grids in the bulk flow region, where turbulent
dissipation or Kolmogorov length scales become larger. Computational optimization, particularly
in the presence of moment-projections of the forcing terms, is also discussed.

Two canonical bounded flows, viz., fully-developed turbulent channel
flow and 3D driven cavity flow have been simulated using this
approach for shear Reynolds number of $183.6$ and Reynolds number
based on cavity side length of $12,000$ respectively. The structure of
turbulent flow given in terms of turbulence statistics, including mean
velocity and components of root-mean-square (rms) velocity and
vorticity fluctuations and Reynolds stress are in good agreement
with prior DNS and experimental data. The computed rms pressure fluctuations
are found to be somewhat over-predicted in comparison with DNS data, which is based
on the solution of incompressible Navier--Stokes equations. It is
thought this may due to the kinetic nature of the GLBE approach, which
is inherently weakly compressible. In the case of 3D cavity flow, the
GLBE is able to capture turbulent velocity fluctuations and Reynolds
stresses generated by cavity walls, which are in reasonably good
agreement with prior data.

With regard to numerical stability, it is found that by separating
various relaxation times, the GLBE is able to maintain solution fidelity, while
the SRT-LBE solution can exhibit spurious effects on velocity fluctuations in
the near-wall region, particularly in the wall normal component, on relatively
coarser grids in turbulent channel flow simulations. The GLBE is found to be
superior in maintaining numerical stability at higher Reynolds
number $\mathrm{Re}$ for 3D cavity flows, with the maximum attainable $\mathrm{Re}$
several times that for the SRT-LBE, depending on the resolution and
shear viscosity of the fluid. Moreover, parallel implementation of the GLBE
approach is able to maintain near-linear scalability in performance
for over a thousand processors on a large parallel cluster.

The GLBE with forcing term appears to be a reliable approach for LES of
wall-bounded turbulent flows. It is expected that further
improvements can be achieved by introducing more advanced SGS models
based on dynamic procedures~\cite{germano91,zhang93,salvetti95} in the
LBM~\cite{premnath08a}.

%\begin{figure}
%\includegraphics[width = 120mm,viewport=60 90 700
%520,clip]{comparemodels}% Here is how to import EPS art
%\caption{\label{fig:comparemodels} Comparison of the components of
%root-mean-square (rms) velocity fluctuations normalized by the wall
%shear velocity for fully-developed turbulent channel flow with a
%rigid free-surface at the top for $Re_{*} = 180$ obtained by GLBE or
%MRT-LBE (dashed lines with open symbols) and BGK-LBE or SRT-LBE
%(solid line with filled symbols).}
%\end{figure}
%%%%%% FIGURE %%%%%

%\begin{figure}
%\includegraphics[width = 120mm,viewport=30 65 720
%570,clip]{parspeedup}% Here is how to import EPS art
%\caption{\label{fig:parallelspeedup} Parallel performance of GLBE
%for LES of fully-developed turbulent channel flow at $Re_{*} =
%183.6$ on IBM p690 cluster at NCSA.}
%\end{figure}
%%%%%% FIGURE %%%%%

\begin{acknowledgments}
This work was performed under the auspices of the National
Aeronautics and Space Administration (NASA) under Contract
Nos.~NNL06AA34P and~NNL07AA04C and U.S. Department of Energy (DOE)
under Grant No.~DE-FG02-03ER83715. Computational resources were
provided by the National Center for Supercomputing Applications
(NCSA) under Award CTS 060027 and the Office of Science of DOE
under Contract DE-AC03-76SF00098.
\end{acknowledgments}

\appendix
\section{\label{app:momentcomponents} Components of Moments, Equilibrium Moments and Moment-Projections of Forcing Terms
for the D3Q19 Lattice}
The components of the various elements in the moments are as
follows~\cite{dhumieres02}: $\widehat{f}_0 = \rho, \widehat{f}_1= e,
\widehat{f}_2 = e^2, \widehat{f}_3 = j_x,\widehat{f}_4 =
q_x,\widehat{f}_5 = j_y, \widehat{f}_6 = q_y, \widehat{f}_7 = j_z,
\widehat{f}_8 = q_z, \widehat{f}_9 = 3p_{xx},\widehat{f}_{10} =
3\pi_{xx},\widehat{f}_{11} = p_{ww},\widehat{f}_{12} =
\pi_{ww},\widehat{f}_{13} = p_{xy},\widehat{f}_{14} =
p_{yz},\widehat{f}_{15} = p_{xz},\widehat{f}_{16} =
m_x,\widehat{f}_{17} = m_y,\widehat{f}_{18} = m_z$.

Here, $\rho$ is the density, $e$ and $e^2$ represent kinetic energy
that is independent of density and square of energy, respectively;
$j_x$, $j_y$ and $j_z$ are the components of the momentum, i.e. $j_x
= \rho u_x$, $j_y = \rho u_y$, $j_z = \rho u_z$, $q_x$, $q_y$, $q_z$
are the components of the energy flux, and $p_{xx}$, $p_{xy}$,
$p_{yz}$ and $p_{xz}$ are the components of the symmetric traceless
viscous stress tensor. The other two normal components of the
viscous stress tensor, $p_{yy}$ and $p_{zz}$, can be constructed
from $p_{xx}$ and $p_{ww}$, where $p_{ww} = p_{yy} - p_{zz}$. Other
moments include $\pi_{xx}$, $\pi_{ww}$, $m_x$, $m_y$ and $m_z$. The
first two of these moments have the same symmetry as the diagonal
part of the traceless viscous tensor $p_{ij}$, while the last three
vectors are parts of a third rank tensor, with the symmetry of
$j_kp_{mn}$.

The components of the equilibrium moments for the D3Q19 lattice are
as follows:~\cite{dhumieres02}: $\widehat{f}_0^{eq} = \rho,
\widehat{f}_1^{eq} \equiv
e^{eq}=-11\rho+19\frac{\overrightarrow{j}\cdot\overrightarrow{j}}{\rho},
\widehat{f}_2^{eq} \equiv
e^{2,eq}=3\rho-\frac{11}{2}\frac{\overrightarrow{j}\cdot\overrightarrow{j}}{\rho},
\widehat{f}_3^{eq} = j_x,\widehat{f}_4^{eq} \equiv
q_x^{eq}=-\frac{2}{3}j_x,\widehat{f}_5^{eq} = j_y,
\widehat{f}_6^{eq} \equiv q_y^{eq}=-\frac{2}{3}j_y,
\widehat{f}_7^{eq} = j_z, \widehat{f}_8^{eq} \equiv
q_z^{eq}=-\frac{2}{3}j_z, \widehat{f}_9^{eq} \equiv
3p_{xx}^{eq}=\frac{\left[3j_x^2-\overrightarrow{j}\cdot\overrightarrow{j}
\right]}{\rho},\widehat{f}_{10}^{eq} \equiv
3\pi_{xx}^{eq}=3\left(-\frac{1}{2}p_{xx}^{eq}
\right),\widehat{f}_{11}^{eq} \equiv
p_{ww}^{eq}=\frac{\left[j_y^2-j_z^2
\right]}{\rho},\widehat{f}_{12}^{eq} \equiv
\pi_{ww}^{eq}=-\frac{1}{2}p_{ww}^{eq},\widehat{f}_{13}^{eq} \equiv
p_{xy}^{eq}=\frac{j_xj_y}{\rho},\widehat{f}_{14}^{eq} \equiv
p_{yz}^{eq}=\frac{j_yj_z}{\rho},\widehat{f}_{15}^{eq} \equiv
p_{xz}^{eq}=\frac{j_xj_z}{\rho},\widehat{f}_{16}^{eq} =
0,\widehat{f}_{17}^{eq} = 0,\widehat{f}_{18}^{eq} = 0$.

The components of the source terms in moment space can be obtained
by multiplying the transformation matrix with Eq.~(\ref{eq:simpleforce}).
The final expressions are as follows:

$ \widehat{S}_0 =   0, \qquad \widehat{S}_1   =
38(F_xu_x+F_yu_y+F_zu_z), \qquad  \widehat{S}_2 =
-11(F_xu_x+F_yu_y+F_zu_z), $

$ \widehat{S}_3= F_x, \qquad \widehat{S}_4 =-\frac{2}{3}F_x, \qquad
\widehat{S}_5=F_y,\qquad \widehat{S}_6 =-\frac{2}{3}F_y, \qquad
\widehat{S}_7=F_z,\qquad \widehat{S}_8 =   -\frac{2}{3}F_z,$

$ \widehat{S}_9   = 2(2F_xu_x-F_yu_y-F_zu_z), \qquad
\widehat{S}_{10} = -(2F_xu_x-F_yu_y-F_zu_z),$

$ \widehat{S}_{11} = 2(F_yu_y-F_zu_z),\qquad \widehat{S}_{12} =
-(F_yu_y-F_zu_z),\qquad \widehat{S}_{13} = (F_xu_y+F_yu_x), $

$ \widehat{S}_{14} = (F_yu_z+F_zu_y), \qquad \widehat{S}_{15} = (F_xu_z+F_zu_x),\qquad \widehat{S}_{16} =   0,
\qquad \widehat{S}_{17} =   0,\qquad \widehat{S}_{18} =   0$.

The self-consistency of the moment projections of source terms is
evident. For e.g., $\widehat{S}_3$, $\widehat{S}_5$ and
$\widehat{S}_7$ provide Cartesian components of body forces on the
moments corresponding to the components of momentum (mass flux),
$\widehat{S}_1$ provides the work due to forces on the moment
corresponding to kinetic energy, etc.

\section{\label{app:strainrate} Strain Rate Tensor using Non-equilibrium Moments in the GLBE with Forcing Term}
In this section, we will present a brief derivation of the strain rate
tensor in terms of the non-equilibrium moments of the GLBE with
forcing term by applying a Chapman--Enskog analysis. The results that follow
are generalizations of those presented by Yu \emph{et~al}.~\cite{yu06} to
include forcing terms representing non-uniform forces. First, the
left hand side of the GLBE is simplified by applying Taylor
series, which results in the following:
\begin{equation}
\delta_t \mathbb{D}_t \mathbf{\widehat{f}} \approx
 -\mathcal{T}^{-1}\widehat{\Lambda}\left( \mathbf{\widehat{f}} -
\mathbf{\widehat{f}}^{eq}\right)+\mathcal{T}^{-1}\left(
\mathrm{I}-\frac{1}{2}\widehat{\Lambda} \right)\mathbf{\widehat{S}}\delta_t,
\label{eq:glbe_taylor}
\end{equation}
where
$
\mathbb{D}_t = diag\left(\partial_{t_0},
\partial_{t_0}+\overrightarrow{e_{1}}\cdot \overrightarrow{\nabla},\ldots, \partial_{t_0}+\overrightarrow{e_{18}}\cdot
\overrightarrow{\nabla}\right) $. Now applying the Chapman--Enskog
expansion
\begin{equation}
\widehat{\mathbf{f}}\approx\widehat{\mathbf{f}}^{eq}+\widehat{\mathbf{f}}^{(1)}\delta_t
\end{equation}
to Eq.~(\ref{eq:glbe_taylor}), and truncating terms of order
$O(\delta_t^2)$ or higher, we get
\begin{equation}
\widehat{\mathbb{D}_t}\widehat{\mathbf{f}}^{eq}=-\widehat{\Lambda}\widehat{\mathbf{f}}^{(1)}+
\left(\mathrm{I} - \frac{1}{2}\widehat{\Lambda} \right)\widehat{\mathbf{S}},\label{eq:glbe_firstorder}
\end{equation}
where
$
\widehat{\mathbb{D}_t} = \mathcal{T} \mathbb{D}_t
\mathcal{T}^{-1}=diag\left(\partial_{t_0},
\partial_{t_0}+\widehat{\Xi}_{1i}\partial_i,\ldots, \partial_{t_0}+\widehat{\Xi}_{18i} \partial_i
\right) $, in which and henceforth summation of repeated indices is
assumed. Eq.~(\ref{eq:glbe_firstorder}) can be rewritten in terms of
non-equilibrium moments $\widehat{\mathbf{f}}^{(1)}$ as
\begin{equation}
\widehat{\mathbf{f}}^{(1)}\approx-\widehat{\Lambda}^{-1}\left(
\partial_{t_0}+ \widehat{\Xi}_{1i} \partial_i
\right)\widehat{\mathbf{f}}^{eq}+ \left( \widehat{\Lambda} -
\frac{1}{2}\mathrm{I} \right)\widehat{\mathbf{S}},\label{eq:non_eqm_moments}
\end{equation}
where $ \widehat{\Xi}_{\alpha i} = \mathcal{T} e_{\alpha i}
\mathcal{T}^{-1} $.

Now substituting the expressions for the equilibrium moments
$\widehat{\mathbf{f}}^{eq}$ and the source terms
$\widehat{\mathbf{S}}$ in Eq. (\ref{eq:non_eqm_moments}), we
simplify the expressions for the components of the non-equilibrium
moments. Some such components of interest are as follows:
\begin{equation}
\widehat{f}^{(1)}_{1}\equiv e^{(1)}=-\frac{1}{s_{1}} \left\{
\partial_{t_0}\left( -11\rho+19\frac{j_kj_k}{\rho} \right)+
\frac{5}{3}\partial_k j_k \right\}+
\left(\frac{1}{s_{1}}-\frac{1}{2}\right)\widehat{S}_{1},
\end{equation}
\begin{equation}
\widehat{f}^{(1)}_{9}\equiv 3p_{xx}^{(1)}=-\frac{1}{s_{9}} \left\{
\partial_{t_0}\left( \frac{j_y^2- j_z^2}{\rho} \right)+
\frac{2}{3}\left(2\partial_x j_x-\partial_y j_y-\partial_z j_z
\right) \right\}+
\left(\frac{1}{s_{9}}-\frac{1}{2}\right)\widehat{S}_{9},
\end{equation}
\begin{equation}
\widehat{f}^{(1)}_{11}\equiv p_{ww}^{(1)}=-\frac{1}{s_{11}} \left\{
\partial_{t_0}\left( \frac{j_y^2- j_z^2}{\rho} \right)+
\frac{2}{3}\left(\partial_y j_y-\partial_z j_z \right) \right\}+
\left(\frac{1}{s_{11}}-\frac{1}{2}\right)\widehat{S}_{11},
\end{equation}
\begin{equation}
\widehat{f}^{(1)}_{13}\equiv p_{xy}^{(1)}=-\frac{1}{s_{13}} \left\{
\partial_{t_0}\left( \frac{j_x j_y}{\rho} \right)+
\frac{1}{3}\left(\partial_x j_y+\partial_y j_x \right) \right\}+
\left(\frac{1}{s_{13}}-\frac{1}{2}\right)\widehat{S}_{13},
\end{equation}
\begin{equation}
\widehat{f}^{(1)}_{14}\equiv p_{yz}^{(1)}=-\frac{1}{s_{14}} \left\{
\partial_{t_0}\left( \frac{j_y j_z}{\rho} \right)+
\frac{1}{3}\left(\partial_y j_z+\partial_z j_y \right) \right\}+
\left(\frac{1}{s_{14}}-\frac{1}{2}\right)\widehat{S}_{14},
\end{equation}
\begin{equation}
\widehat{f}^{(1)}_{15}\equiv p_{xz}^{(1)}=-\frac{1}{s_{15}}
\left\{
\partial_{t_0}\left( \frac{j_x j_z}{\rho} \right)+
\frac{1}{3}\left(\partial_x j_z+\partial_z j_x \right) \right\}+
\left(\frac{1}{s_{15}}-\frac{1}{2}\right)\widehat{S}_{15}.
\end{equation}.

For further simplification, we invoke the following approximations:
$
\partial_{t_0}\left( \frac{j_x^2}{\rho} \right)\simeq 2u_x F_x,
\quad
\partial_{t_0}\left( \frac{j_y^2}{\rho} \right)\simeq 2u_y F_y,
\quad
\partial_{t_0}\left( \frac{j_z^2}{\rho} \right)\simeq 2u_z F_z,
$ $
\partial_{t_0}\left( \frac{j_x j_y}{\rho} \right)\simeq u_x F_y + u_y F_x,
\quad
\partial_{t_0}\left( \frac{j_y j_z}{\rho} \right)\simeq u_y F_z + u_z F_y,
\quad
\partial_{t_0}\left( \frac{j_x j_z}{\rho} \right)\simeq u_x F_z + u_z F_x,
$ and $
\partial_{t_0}\rho =-\partial_k j_k,
$ which result in
\begin{equation}
\widehat{f}^{(1)}_{1}=-\frac{38}{3}\frac{1}{s_1}\partial_k j_k
-\frac{1}{2}\widehat{S}_1,
\end{equation}
\begin{equation}
\widehat{f}^{(1)}_{9}\approx-\frac{2}{3}\frac{1}{s_{9}}\left(3\partial_x
j_x-\partial_k j_k \right) -\frac{1}{2}\widehat{S}_{9},
\end{equation}
\begin{equation}
\widehat{f}^{(1)}_{11}\approx-\frac{2}{3}\frac{1}{s_{11}}\left(\partial_y
j_y-\partial_z j_z \right) -\frac{1}{2}\widehat{S}_{11},
\end{equation}
\begin{equation}
\widehat{f}^{(1)}_{13}\approx-\frac{1}{3}\frac{1}{s_{13}}\left(\partial_x
j_y+\partial_y j_x \right) -\frac{1}{2}\widehat{S}_{13},
\end{equation}
\begin{equation}
\widehat{f}^{(1)}_{14}\approx-\frac{1}{3}\frac{1}{s_{14}}\left(\partial_y
j_z+\partial_z j_y \right) -\frac{1}{2}\widehat{S}_{14},
\end{equation}
\begin{equation}
\widehat{f}^{(1)}_{15}\approx-\frac{1}{3}\frac{1}{s_{15}}\left(\partial_x
j_z+\partial_z j_x \right) -\frac{1}{2}\widehat{S}_{15}.
\end{equation}

It follows from the above that the components of the strain rate
tensor can be written explicitly in terms of non-equilibrium moments
as
\begin{eqnarray}
S_{xx}&\approx&-\frac{1}{38\rho} \left[
s_{1}\widehat{h}_{1}^{(neq)}+19s_{9}\widehat{h}_{9}^{(neq)}  \right], \label{eq:sxx}\\
S_{yy}&\approx&-\frac{1}{76\rho} \left[
2s_{1}\widehat{h}_{1}^{(neq)}-19\left(s_{9}\widehat{h}_{9}^{(neq)}-3s_{11}\widehat{h}_{11}^{(neq)}\right)
\right],\label{eq:syy}\\
S_{zz}&\approx&-\frac{1}{76\rho} \left[
2s_{1}\widehat{h}_{1}^{(neq)}-19\left(s_{9}\widehat{h}_{9}^{(neq)}+3s_{11}\widehat{h}_{11}^{(neq)}\right)
\right],\label{eq:szz}\\
S_{xy}&\approx&-\frac{3}{2\rho}
s_{13}\widehat{h}_{13}^{(neq)}, \label{eq:sxy}\\
S_{yz}&\approx&-\frac{3}{2\rho}
s_{14}\widehat{h}_{14}^{(neq)}, \label{eq:syz}\\
S_{xz}&\approx&-\frac{3}{2\rho}
s_{15}\widehat{h}_{15}^{(neq)},\label{eq:sxz}
\end{eqnarray}
where
\begin{equation}
\widehat{h}_{\alpha}^{(neq)}=\widehat{f}_{\alpha}-\widehat{f}_{\alpha}^{eq}+\frac{1}{2}\widehat{S}_{\alpha},\qquad
\alpha \in \{1,9,11,13,14,15\}
\end{equation}
Here, the components of the source term $\widehat{S}_{\alpha}$ can
be obtained from Appendix~\ref{app:momentcomponents}. The form of $S_{ij}$ turns out
to be very similar to that obtained by Yu \emph{et~al}.~\cite{yu06}, except for the
expression ${h}_{\alpha}^{(neq)}$, which contains the additional contribution $\frac{1}{2}\widehat{S}_{\alpha}$
that provides the effect of the forcing term. The procedure discussed here, however, is general,
and can be readily employed for
deriving the expressions for strain rate tensor for other lattice velocity models in the presence of forcing terms.
The magnitude of the strain rate $|S|$ used in turbulence models can then be
obtained from Eqs. (\ref{eq:sxx})--(\ref{eq:sxz}) as $|S|=\sqrt{2S_{ij}S_{ij}}=\sqrt{2(S_{xx}^2+S_{yy}^2+S_{zz}^2+2(S_{xy}^2+S_{yz}^2+S_{xz}^2))}$.
To clarify the notations employed, we again note that $\widehat{S}_{\alpha}$
represents the source terms in moment space, $s_{\alpha}$
corresponds to the relaxation times in the collision term, and
$S_{ij}$ is the strain rate tensor.

\newpage %Just because of unusual number of tables stacked at end
%\bibliography{channel}% Produces the bibliography via BibTeX.
% Replace bibliography with the .bbl file contents

\end{document}